\documentstyle[eqsecnum,rotating,epsfig,aps]{revtex}
 \begin{document}
\newcommand{\str}{\rule{0ex}{2.7ex}}  
%
\newcommand{\pr}{Phys.\ Rev.\ }
\newcommand{\prp}{Phys.\ Rep.\  }
\newcommand{\np}{Nucl.\ Phys.\ B }
\newcommand{\zp}{Z.\ Phys.\ C }
\newcommand{\beq}{\begin{equation}}
\newcommand{\eeq}{\end{equation}}
\newcommand{\bea}{\begin{eqnarray}}
\newcommand{\eea}{\end{eqnarray}}
%
\draft
\preprint{}
%
\newbox\hdbox%
\newcount\hdrows%
\newcount\multispancount%
\newcount\ncase%
\newcount\ncols
\newcount\nrows%
\newcount\nspan%
\newcount\ntemp%
\newdimen\hdsize%
\newdimen\newhdsize%
\newdimen\parasize%
\newdimen\spreadwidth%
\newdimen\thicksize%
\newdimen\thinsize%
\newdimen\tablewidth%
\newif\ifcentertables%
\newif\ifendsize%
\newif\iffirstrow%
\newif\iftableinfo%
\newtoks\dbt%
\newtoks\hdtks%
\newtoks\savetks%
\newtoks\tableLETtokens%
\newtoks\tabletokens%
\newtoks\widthspec%
%
%
\immediate\write15{%
CP SMSG GJMSINK TEXTABLE --> TABLE MACROS V. 851121 JOB = \jobname%
}%
%
%
\tableinfotrue%
\catcode`\@=11
\def\out#1{\immediate\write16{#1}}
%
%
\def\tstrut{\vrule height3.1ex depth1.2ex width0pt}%
\def\and{\char`\&}
\def\tablerule{\noalign{\hrule height\thinsize depth0pt}}%
\thicksize=1.5pt
\thinsize=0.6pt
\def\thickrule{\noalign{\hrule height\thicksize depth0pt}}%
\def\hrulefill{\leaders\hrule\hfill}%
\def\bigrulefill{\leaders\hrule height\thicksize depth0pt \hfill}%
\def\ctr#1{\hfil\ #1\hfil}%
\def\altctr#1{\hfil #1\hfil}%
\def\vctr#1{\hfil\vbox to0pt{\vss\hbox{#1}\vss}\hfil}%
%
%
\tablewidth=-\maxdimen%
\spreadwidth=-\maxdimen%
\def\tabskipglue{0pt plus 1fil minus 1fil}%
%
%
\centertablestrue%
\def\centeredtables{%
   \centertablestrue%
}%
\def\noncenteredtables{%
   \centertablesfalse%
}%
%
%
\parasize=4in%
\long\def\para#1{
   {%
      \vtop{%
         \hsize=\parasize%
         \baselineskip14pt%
         \lineskip1pt%
         \lineskiplimit1pt%
         \noindent #1%
         \vrule width0pt depth6pt%
      }%
   }%
}%
\gdef\ARGS{########}
\gdef\headerARGS{####}
\def\@mpersand{&}
{\catcode`\|=13
\gdef\letbarzero{\let|0}
\gdef\letbartab{\def|{&&}}%
\gdef\letvbbar{\let\vb|}%
}
{\catcode`\&=4
\def\ampskip{&\omit\hfil&}
\catcode`\&=13
\let&0
\xdef\letampskip{\def&{\ampskip}}%
\gdef\letnovbamp{\let\novb&\let\tab&}
}
\def\begintable{
   \begingroup%
   \catcode`\|=13\letbartab\letvbbar%
   \catcode`\&=13\letampskip\letnovbamp%
   \def\multispan##1{
      \omit \mscount##1%
      \multiply\mscount\tw@\advance\mscount\m@ne%
      \loop\ifnum\mscount>\@ne \sp@n\repeat%
   }
   \def\|{%
      &\omit\widevline&%
   }%
   \ruledtable
}
\long\def\ruledtable#1\endtable{%
%
%
%
   \offinterlineskip
   \tabskip 0pt
   \def\widevline{\vrule width\thicksize}
   \def\endrow{\@mpersand\omit\hfil\crnorm\@mpersand}%
   \def\crthick{\@mpersand\crnorm\thickrule\@mpersand}%
   \def\crthickneg##1{\@mpersand\crnorm\thickrule
          \noalign{{\skip0=##1\vskip-\skip0}}\@mpersand}%
   \def\crnorule{\@mpersand\crnorm\@mpersand}%
   \def\crnoruleneg##1{\@mpersand\crnorm
          \noalign{{\skip0=##1\vskip-\skip0}}\@mpersand}%
   \let\nr=\crnorule
   \def\endtable{\@mpersand\crnorm\thickrule}%
   \let\crnorm=\cr
%
%
   \edef\cr{\@mpersand\crnorm\tablerule\@mpersand}%
   \def\crneg##1{\@mpersand\crnorm\tablerule
          \noalign{{\skip0=##1\vskip-\skip0}}\@mpersand}%
   \let\ctneg=\crthickneg
   \let\nrneg=\crnoruleneg
   \the\tableLETtokens
%
%
   \tabletokens={&#1}
%
%
   \countROWS\tabletokens\into\nrows%
   \countCOLS\tabletokens\into\ncols%
%
%
   \advance\ncols by -1%
   \divide\ncols by 2%
   \advance\nrows by 1%
%
%
   \iftableinfo %
      \immediate\write16{[Nrows=\the\nrows, Ncols=\the\ncols]}%
   \fi%
%
%
   \ifcentertables
      \ifhmode \par\fi
      \hbox to \hsize{
      \hss
   \else %
      \hbox{%
   \fi
      \vbox{%
         \makePREAMBLE{\the\ncols}
         \edef\next{\preamble}
         \let\preamble=\next
         \makeTABLE{\preamble}{\tabletokens}
      }
      \ifcentertables \hss}\else }\fi
   \endgroup
   \tablewidth=-\maxdimen
   \spreadwidth=-\maxdimen
}
\def\makeTABLE#1#2{
   {
   \let\ifmath0
   \let\header0
   \let\multispan0
%
%
   \ncase=0%
   \ifdim\tablewidth>-\maxdimen \ncase=1\fi%
   \ifdim\spreadwidth>-\maxdimen \ncase=2\fi%
   \relax
%
   \ifcase\ncase %
      \widthspec={}%
   \or %
      \widthspec=\expandafter{\expandafter t\expandafter o%
                 \the\tablewidth}%
   \else %
      \widthspec=\expandafter{\expandafter s\expandafter p\expandafter r%
                 \expandafter e\expandafter a\expandafter d%
                 \the\spreadwidth}%
   \fi %
   \xdef\next{
      \halign\the\widthspec{%
      #1
      \noalign{\hrule height\thicksize depth0pt}
      \the#2\endtable
%
      }
   }
   }
   \next
}
\def\makePREAMBLE#1{
   \ncols=#1
   \begingroup
   \let\ARGS=0
   \edef\xtp{\widevline\ARGS\tabskip\tabskipglue%
   &\ctr{\ARGS}\tstrut}
   \advance\ncols by -1
   \loop
      \ifnum\ncols>0 %
      \advance\ncols by -1%
      \edef\xtp{\xtp&\vrule width\thinsize\ARGS&\ctr{\ARGS}}%
   \repeat
   \xdef\preamble{\xtp&\widevline\ARGS\tabskip0pt%
   \crnorm}
   \endgroup
}
\def\countROWS#1\into#2{
   \let\countREGISTER=#2%
   \countREGISTER=0%
   \expandafter\ROWcount\the#1\endcount%
}%
\def\ROWcount{%
   \afterassignment\subROWcount\let\next= %
}%
\def\subROWcount{%
   \ifx\next\endcount %
      \let\next=\relax%
   \else%
      \ncase=0%
      \ifx\next\cr %
         \global\advance\countREGISTER by 1%
         \ncase=0%
      \fi%
      \ifx\next\endrow %
         \global\advance\countREGISTER by 1%
         \ncase=0%
      \fi%
      \ifx\next\crthick %
         \global\advance\countREGISTER by 1%
         \ncase=0%
      \fi%
      \ifx\next\crnorule %
         \global\advance\countREGISTER by 1%
         \ncase=0%
      \fi%
      \ifx\next\crthickneg %
         \global\advance\countREGISTER by 1%
         \ncase=0%
      \fi%
      \ifx\next\crnoruleneg %
         \global\advance\countREGISTER by 1%
         \ncase=0%
      \fi%
      \ifx\next\crneg %
         \global\advance\countREGISTER by 1%
         \ncase=0%
      \fi%
      \ifx\next\header %
         \ncase=1%
      \fi%
      \relax%
      \ifcase\ncase %
         \let\next\ROWcount%
      \or %
         \let\next\argROWskip%
      \else %
      \fi%
   \fi%
   \next%
}
\def\counthdROWS#1\into#2{%
\dvr{10}%
   \let\countREGISTER=#2%
   \countREGISTER=0%
\dvr{11}%
\dvr{13}%
   \expandafter\hdROWcount\the#1\endcount%
\dvr{12}%
}%
\def\hdROWcount{%
   \afterassignment\subhdROWcount\let\next= %
}%
\def\subhdROWcount{%
   \ifx\next\endcount %
      \let\next=\relax%
   \else%
      \ncase=0%
      \ifx\next\cr %
         \global\advance\countREGISTER by 1%
         \ncase=0%
      \fi%
      \ifx\next\endrow %
         \global\advance\countREGISTER by 1%
         \ncase=0%
      \fi%
      \ifx\next\crthick %
         \global\advance\countREGISTER by 1%
         \ncase=0%
      \fi%
      \ifx\next\crnorule %
         \global\advance\countREGISTER by 1%
         \ncase=0%
      \fi%
      \ifx\next\header %
         \ncase=1%
      \fi%
\relax%
      \ifcase\ncase %
         \let\next\hdROWcount%
      \or%
         \let\next\arghdROWskip%
      \else %
      \fi%
   \fi%
   \next%
}%
{\catcode`\|=13\letbartab
\gdef\countCOLS#1\into#2{%
   \let\countREGISTER=#2%
   \global\countREGISTER=0%
   \global\multispancount=0%
   \global\firstrowtrue
   \expandafter\COLcount\the#1\endcount%
   \global\advance\countREGISTER by 3%
   \global\advance\countREGISTER by -\multispancount
}%
\gdef\COLcount{%
   \afterassignment\subCOLcount\let\next= %
}%
{\catcode`\&=13%
\gdef\subCOLcount{%
   \ifx\next\endcount %
      \let\next=\relax%
   \else%
      \ncase=0%
      \iffirstrow
         \ifx\next& %
            \global\advance\countREGISTER by 2%
            \ncase=0%
         \fi%
         \ifx\next\span %
            \global\advance\countREGISTER by 1%
            \ncase=0%
         \fi%
         \ifx\next| %
            \global\advance\countREGISTER by 2%
            \ncase=0%
         \fi
         \ifx\next\|
            \global\advance\countREGISTER by 2%
            \ncase=0%
         \fi
         \ifx\next\multispan
            \ncase=1%
            \global\advance\multispancount by 1%
         \fi
         \ifx\next\header
            \ncase=2%
         \fi
         \ifx\next\cr       \global\firstrowfalse \fi
         \ifx\next\endrow   \global\firstrowfalse \fi
         \ifx\next\crthick  \global\firstrowfalse \fi
         \ifx\next\crnorule \global\firstrowfalse \fi
         \ifx\next\crnoruleneg \global\firstrowfalse \fi
         \ifx\next\crthickneg  \global\firstrowfalse \fi
         \ifx\next\crneg       \global\firstrowfalse \fi
      \fi
\relax
      \ifcase\ncase %
         \let\next\COLcount%
      \or %
         \let\next\spancount%
      \or %
         \let\next\argCOLskip%
      \else %
      \fi %
   \fi%
   \next%
}%
\gdef\argROWskip#1{%
   \let\next\ROWcount \next%
}
\gdef\arghdROWskip#1{%
   \let\next\ROWcount \next%
}
\gdef\argCOLskip#1{%
   \let\next\COLcount \next%
}
}
}
\def\spancount#1{
   \nspan=#1\multiply\nspan by 2\advance\nspan by -1%
   \global\advance \countREGISTER by \nspan
   \let\next\COLcount \next}%
\def\dvr#1{\relax}%
\def\header#1{%
\dvr{1}{\let\cr=\@mpersand%
\hdtks={#1}%
\counthdROWS\hdtks\into\hdrows%
\advance\hdrows by 1%
\ifnum\hdrows=0 \hdrows=1 \fi%
\dvr{5}\makehdPREAMBLE{\the\hdrows}%
\dvr{6}\getHDdimen{#1}%
{\parindent=0pt\hsize=\hdsize{\let\ifmath0%
\xdef\next{\valign{\headerpreamble #1\crnorm}}}\dvr{7}\next\dvr{8}%
}%
}\dvr{2}}
\def\makehdPREAMBLE#1{
\dvr{3}%
\hdrows=#1
{
\let\headerARGS=0%
\let\cr=\crnorm%
\edef\xtp{\vfil\hfil\hbox{\headerARGS}\hfil\vfil}%
\advance\hdrows by -1
\loop
\ifnum\hdrows>0%
\advance\hdrows by -1%
\edef\xtp{\xtp&\vfil\hfil\hbox{\headerARGS}\hfil\vfil}%
\repeat%
\xdef\headerpreamble{\xtp\crcr}%
}
\dvr{4}}
\def\getHDdimen#1{%
\hdsize=0pt%
\getsize#1\cr\end\cr%
}
\def\getsize#1\cr{%
\endsizefalse\savetks={#1}%
\expandafter\lookend\the\savetks\cr%
\relax \ifendsize \let\next\relax \else%
\setbox\hdbox=\hbox{#1}\newhdsize=1.0\wd\hdbox%
\ifdim\newhdsize>\hdsize \hdsize=\newhdsize \fi%
\let\next\getsize \fi%
\next%
}%
\def\lookend{\afterassignment\sublookend\let\looknext= }%
\def\sublookend{\relax%
\ifx\looknext\cr %
\let\looknext\relax \else %
   \relax
   \ifx\looknext\end \global\endsizetrue \fi%
   \let\looknext=\lookend%
    \fi \looknext%
}%
%
%
\def\tablelet#1{%
   \tableLETtokens=\expandafter{\the\tableLETtokens #1}%
}%
\catcode`\@=12
%

\title{ 
     New features in	DPMJET    version  II.5      
}

 \author{  J.~Ranft }
  \address{
 Physics Dept. Universit\"at Siegen, D--57068 Siegen, Germany,
 e--mail: Johannes.Ranft@cern.ch}
\date{\today}


\maketitle


\vspace{3mm}
%

\begin{abstract}
DPMJET is a Monte Carlo model for
        sampling of hadron--hadron, hadron-nucleus,
           nucleus-nucleus  and neutrino--nucleus interactions 
          at accelerator and Cosmic Ray energies 
	according to the two--component Dual Parton Model

Here we describe new features in version DPMJET--II.5:
Implementation of new DPM diagrams for an improved description
of baryon stopping in nuclear collisions and improvements in the
calculation of Glauber cross sections. The new diagrams allow
two quite different extrapolations of the model to the highest
Cosmic Ray energies. The new version of the model is compared to
experimental data on hadron--hadron, hadron--nucleus and
nucleus--nucleus collisions.
\end{abstract}

\vskip 10mm 
{Siegen preprint SI--99-5}

\section{Introduction}

In the present paper we desribe new features in 
the physics of DPMJET--II.5 and compare
important features of the model to experimental data. The code manual
for DPMJET--II.5 will be given in a separate paper \cite{Ranft99b}.

 The DPMJET--II.2
event generator  was described in detail  before
 \cite{DPMJETII,Ranftsare95,Ranft95b}.
The main topic in switching to DPMJET--II.3/II.4   
  \cite{dpmjet2324} was
 the extension of the model up to energies of $\sqrt s$ =
2000 TeV. 
 The two versions DPMJET--II.3 and DPMJET--II.4 did differ mainly by the
versions of the Lund code JETSET they use. DPMJET--II.3 did use
a JETSET version transformed into DOUBLE PRECISION. DPMJET--II.4
uses 
the 1997 version PYTHIA--6.1 now in DOUBLE PRECISION
in which JETSET is contained. DPMJET--II.5 continues to use
PYTHIA--6.1.
The extension to higher energies 
was done in DPMJET--II.3/4 by calculating the minijet component of
the model using  as default the
 GRV--LO94 parton distributions
\cite{Gluck95a}. 
This is replaced in DPMJET--II.5 
by the more recent version GRV--LO--98 \cite{Gluck98a}.
 These new parton
distributions describe the data measured in the last years at the
HERA Collider.

DPMJET--II is used in Collaboration with Battistoni, Carboni and Forti
\cite{Battistoni99a} for the simulation of the Cosmic Ray cascade in the
HEMAS--DPM code system.
A hadron production model to be used  for simulations at
accelerator and  Cosmic Ray energies
should use all possible information from fixed target
experiments and collider experiments at accelerators. There might
be large  differences on what is considered important:
Best
studied at accelerators is the central region  in hadron--hadron
collisions.
 For studying the Cosmic Ray
cascade, the main interest is in the forward fragmentation
region of hadron--nucleus and nucleus--nucleus collisions.
DPMJET simulates  hadron production in the framework
of the Dual Parton Model with emphasis  as well in the central
as in the fragmentation
region. 

%
Soft multiparticle production characterizing hadronic  
interactions at supercollider  or Cosmic Ray 
energies cannot be understood purely within theoretical 
approaches provided   by perturbative QCD. The
nonperturbative soft component of hadron production, which is
responsible for all of hadron production at low energies is
still acting at higher energies together with the hard component
described by perturbative QCD. 

Using basic ideas of the dual topological unitarization scheme
\cite{Chew76,Chan75}
the Dual Parton Model (DPM) \cite{Capella79} 
(a published review stressing mainly
non-Monte Carlo applications is given in 
Ref.\cite{capdpm}, a nice review, also with emphasis to
Monte Carlo implementations of the DPM--model and to photon induced
reactions  can be found in
the Ph.D Thesis of Ralph Engel \cite{Engelthesis} )
has been very successfully describing soft hadronic processes
together with the gradual transition to hard collisions with
rising energy. 

Observations
like rapidity plateaus and average transverse momenta 
rising with energy,
KNO scaling violation, 
transverse momentum--multiplicity correlations and
{\it minijets} pointed out, 
that soft and hard processes are closely related.
These properties were understood within the two--component Dual Parton
Model
\cite{CTK87,DTUJETPR92,DTUJETZP91,DTUJET92a,DTUJET92b,DTUJETDIFF,DTUJET93}. 

The hard component is introduced applying lowest order
of perturbative hard constituent scattering \cite{CKR77}. 
Single diffraction dissociation
is represented by a triple--pomeron
exchange (high mass single diffraction) and a low mass component.

The Dual Parton model provides a framework not only for the
study of hadron--hadron interactions, but also for the
description of particle production in hadron--nucleus and
nucleus--nucleus collisions at high energies. Within this model
the high energy projectile undergoes a multiple scattering as
formulated in Glaubers approach; particle production is 
realized by the fragmentation of colorless parton--parton chains
constructed from the quark content of the 
interacting hadrons and nuclei.

The first successfull applications
of the Monte Carlo version of the dual parton model (DPM)
to hadron--nucleus \cite{RaRiHA,RaRiHA1}
and nucleus--nucleus \cite{MRR85,RaKK87,RaZPC89} collisions 
also demonstrated
that the cascade of created secondaries in the target (and projectile)
nuclei contributes significantly to particle production
in the target (and projectile) fragmentation regions.
DPMJET uses 
the concept of a formation zone \cite{LaPom53,Sto75}
suppressing in a natural way the cascading of high--energy secondaries.
The  Monte Carlo model 
includes intranuclear cascade processes of the created
secondaries combined with the formation time concept since the
first version of DPMJET--II.
The FZIC (formation zone intranuclear cascade) was also
introduced for the projectile nucleus and in addition the
nuclear evaporation and fragmentation of the residual nucleus
was introduced into DPMJET--II \cite{Ferrari95a,Ferrari96a}.
\\		
In the following we briefly sketch new features  of the model
and mention the most important ingredients;
for a more detailed description of the model as applied in the code
we refer to 
Refs.~\cite{dpmjet2324,Ranft95b,RaRiHA,RaRiHA1,Ranft88a,Moehring91,KMRQM91,KMR92,MRCT93,RER93,CTR1,CTR94,Ferrari95a,Ferrari96a,dpmcharm,DPMBFR94,Battistoni96a}.
The application of DPMJET to neutrino--nucleus interactions is desribed
in \cite{Battistoni98}.
In Section II we will describe new features of 
 the
Dual Parton Model as used in DPMJET--II.5. The most important of
these new features are new diagrams contributing to baryon
stopping and a better calculation of Glauber cross sections.
In Section III we compare the model to data and in Section IV we
discuss the properties of the model at the highest energies. In
both these Sections we present only material obtained with
DPMJET--II.5. 
It is not to be expected that  in the energy region where experimental
data are available the model has changed so
strongly, that comparisons given in former papers with previous
versions of the model would have
become invalid.
%

\centerline{\bf What is new in DPMJET--II.5?}

Here we only summarize the new features
of DPMJET--II.5 against DPMJET--II.3/4 \cite{dpmjet2324} , 
which are described in more
detail in later Sections.

{ \bf Modifications of the Glauber model calculation of nuclear
cross sections.}
For some light nuclei the Woods--Saxon nuclear densities are replaced
by parametrizations in better agreement to data and  measured nuclear
radii are used\cite{Barrett77}  
instead of the nuclear radii according to a
parametrization used before. The new 
option XSECNUC is added to calculate a
table of total, elastic, quasi--elastic and production cross sections
using a modified version of the routine XSGLAU adopted from DTUNUC--II 
  \cite{Roesler96b,Roesler99}.
These changes lead to an improved agreement to measured nuclear cross
sections , especially the p--Air cross sections measured in Cosmic Ray
experiments.

{\bf Elimination of some old features of the code.}
Originally, DPMJET was written in single precision and compiler options
were used on 32 bit word computers to transform the code to DOUBLE
PRECISION. Now DPMJET is written explicitely in DOUBLE PRECISION. A
number of features of the code was changed to get it running on
different UNIX systems, especially HP--unix, DEC--alphas, IBM--unix,
Sun--workstations and  LINUX.

Old features of the code were removed, in particular: the optional
fragmentation of strings according to the BAMJET code
 \cite{BAMJET,BAMJET1}. Only the PYTHIA (JETSET)string fragmentation
 remains in the present DPMJET, we use PYTHIA--6.1 in the DOUBLE
 PRECISION version.  However, the internal particle codes used are
 still the ones from BAMJET, The output particle codes are as before the
ones according to the PDG convention. Some of the older parton structure
 functions are dropped from the code. The chain fusion option, which was
 no longer up to date, has been removed. 

{\bf Single diffraction also implemented in nucleus--nucleus
collisions.}
The present version of the model 
DPMJET--II.5 allows to include or exclude single diffractive events
also in the case of nucleus--nucleus collisions or to sample only single
diffractive events. The diffractive cross section in hadron--nucleus and
nucleus--nucleus collisions is calculated by the Monte Carlo method.

{\bf Implementation of new DPM diagrams for an improved
description of baryon stopping in hadron--nucleus and nucleus--nucleus
collisions.}
 A new striking feature of hadron production in  nuclear
 collisions is the large stopping of the
 participating nucleons (compared to hadron--hadron collisions)
 in hadron--nucleus and nucleus--nucleus collisions 
 \cite{NA35FIN,Alber98}. The {popcorn mechanism} implemented in
 models with independent string fragmentation like the DPM since
 more than 15 years is not sufficient to explain this enhanced
 baryon stopping.
 New DPM--diagrams were proposed by Capella and Kopeliovich
 \cite{Capella96} and investigated in detail by Capella and
 Collaborators
  \cite{Capella99a,Capella99} and by Casado \cite{Casado99}.
 DPMJET--II.5 implements these diagrams and obtains an
 improved agreement to net--baryon distributions in nuclear
 collisions.

{\bf Implementation of the new diagrams allows for two different
extrapolations of the model to Cosmic Ray energies beyond the accelerator
energy range.}
The diagram investigated by Capella and 
Collaborators \cite{Capella99a,Capella99}
uses sea quarks, which in nuclear collisions according to the
Glauber model are needed to implement the multiple collisions in
the nucleus. In the Dual Parton Model in the collider energy
range there are further multiple collisions, again implemented
using sea quarks at the ends of multiple chains, which occur due
to the unitarization procedure. 

We get at high energy enhanced
baryon stopping even in proton--proton collisions 
and a strong modification of the model if also
these sea quarks are used in the new diagrams 
 The  two models obtained with  or without  this
effect result in quite different model predictions at the
highest Cosmic Ray energies.

{\bf Improvements in the sampling and analysis of central
collisions.}
 Older versions of DPMJET were
 also applied to central heavy ion collisions at CERN--SPS, RHIC
 and LHC energies.
The previous version of DPMJET 
 DPMJET--II.3/4 \cite{dpmjet2324},  was mainly implemented with
 the aim to sample minimum bias 
 hadron--nucleus and nucleus--nucleus collisions for
 the simulation of the Cosmic Ray cascade. Heavy targets or
 projectiles like Pb or Au and central collisions are rather
 unimportant for this purpose. 

 Many new, previously unknown features of
 multiparticle production have been found by 
 the experimental heavy ion program at the CERN--SPS.
 It is certainly useful to use  these
 results as a check even if the aim would be just  to simulate
 Cosmic Ray events.
  Important features of these
 results concern in particular strangeness production and the
 baryon stopping discussed already above in central
 nucleus--nucleus collisions.

DPMJET--II.5 does not  desribe all features of the heavy
ion experiments, nevertheless, it has been tuned to work for
central collisions and to describe already quite a number of
features of the
CERN--SPS heavy ion data.

%
\section{Basic physical concepts of the Dual Parton Model}

\subsection{
The two--component Dual Parton Model for hadron--hadron
collisions}

The two component Dual Parton model for hadron--hadron
collisions as applied in DPMJET has been described in detail in
\cite{dpmjet2324,DPMJETII,DTUJETPR92,DTUJET93}. We will not
repeat this here.

The only topic which has to be discussed is the fit of the model
parameters using the new parton structure function GRV--LO--98
\cite{Gluck98a}.

\begin{figure}[thb]
\begin{center}
 \psfig{figure=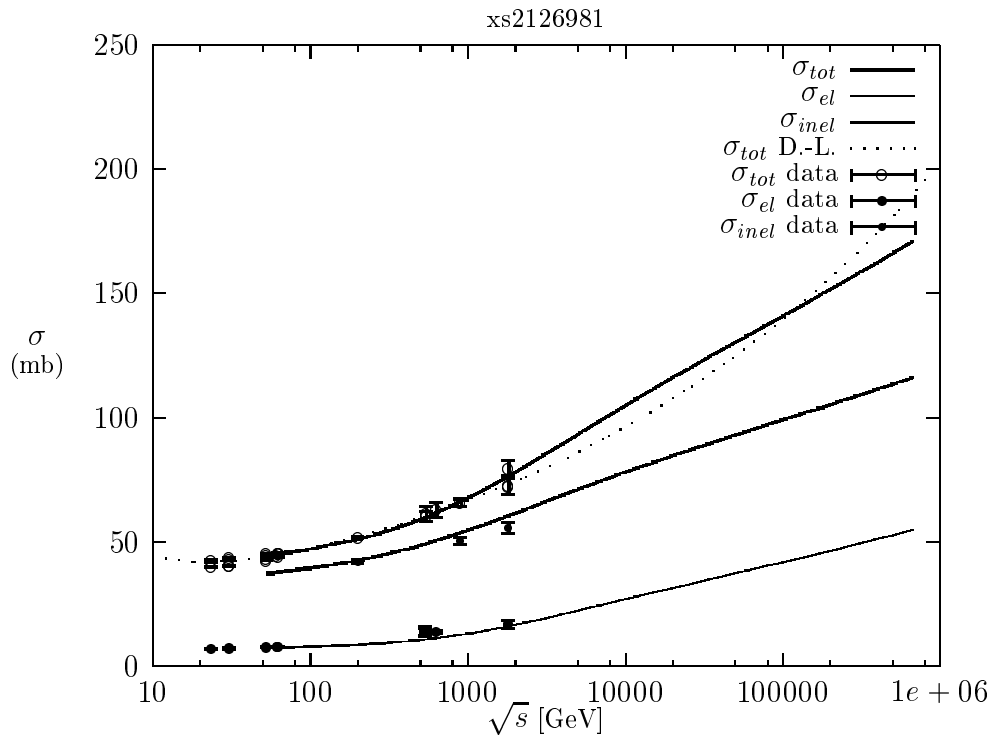}
\end{center}
\vspace*{-3mm}
\caption{Total, inelastic and elastic $p\bar p$ and $pp$ cross
sections as function of the center of mass energy $\sqrt s$. The
model results obtained using the GRV--LO98 parton distributions
\protect\cite{Gluck98a}  
are compared to the Donnachie--Landshoff fit for the total cross
section \protect\cite{Donnachie93} and to data 
\protect\cite{Arnison83,Bozzo84a,Amos85,Bernard87a,Alner86a,Amos90a,Abe93a,Abe94b,Abe94d}
\protect\label{xs2126981}
}
\end{figure}
\begin{figure}[thb]
\begin{center}
 \psfig{figure=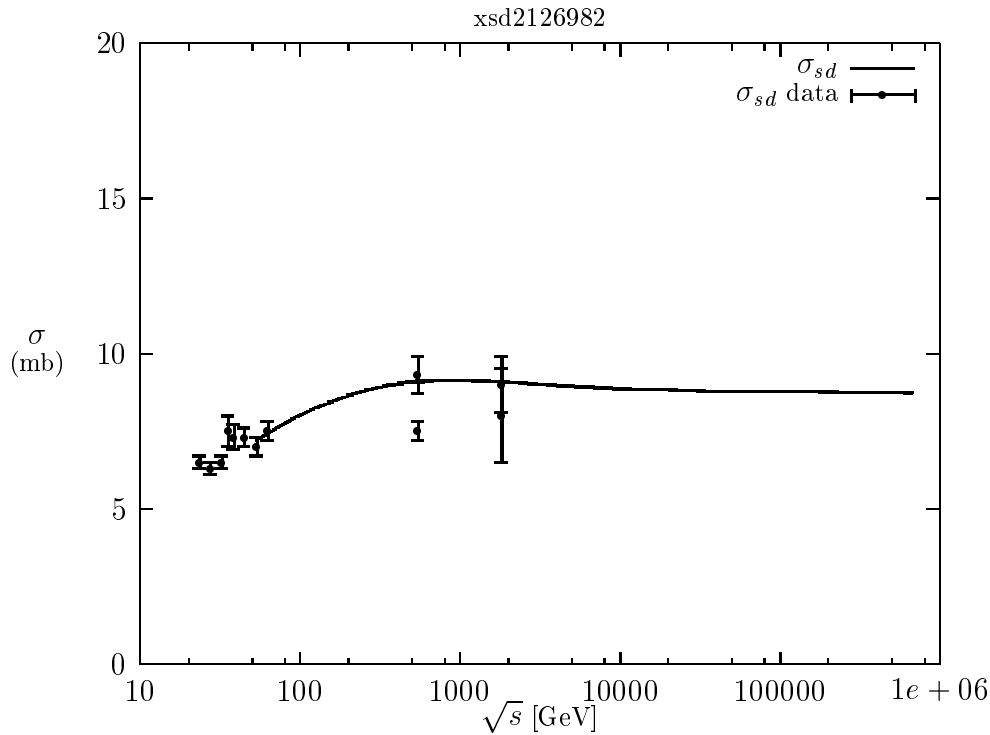}
\end{center}
\vspace*{-3mm}
\caption{Single diffractive $p\bar p$ and $pp$ cross
sections as function of the center of mass energy $\sqrt s$. The
model results obtained using the GRV--LO98 parton distributions
\protect\cite{Gluck98a} are compared to data for \protect$M^2/s = \xi
\leq 0.05$ \protect\cite{Albrow76,Armitage82,Amos93a,Abe94c,Bernard87b}. 
 to data.
\protect\label{xsd2126981}
}
\end{figure}
 
 We  fit  the Pomeron parameters 
 with the energy dependent $p_{\perp}$ cut--off 
 \cite{DTUJET93} for minijets:

\begin{equation}
 p_{\perp thr} = 2.5 + 0.12[\log_{10}(\sqrt s/\sqrt s_0)]^3~~~ 
 {\rm [GeV/c]},~~~ \sqrt s_0 = 50 {\rm GeV}.
\end{equation}

To describe the high energy
particle production we have to determine the free parameters of the
model, i.e. the proton-Pomeron coupling constant $g$, the effective soft
Pomeron intercept $\alpha(0)$, the slope of the Pomeron
trajectory $\alpha^\prime$, the slope parameters $b$ and $b_h$
and the excitation coupling constant
$\lambda$.  This has been done by a global fit to all available data of
total, elastic, inelastic, and single diffractive cross sections in the
energy range from ISR to collider experiments as well as to the
data on the elastic slopes in this energy range.
Since there are some differences
in the hard parton distribution functions at small $x$
values resulting in different hard input cross sections we have
to perform separate fits for each set of
parton distribution functions.
 
We get  good fits using the GRV--LO--98 PDF.
In Table \ref{DTUJET97fit} we give  the parameters obtained in
the fit. The values obtained for $\alpha (0)$ demonstrate again that
the fits result in a supercritical Pomerom.
In Fig. \ref{xs2126981}  we plot the fitted
cross sections  together with 
the data. Furthermore we compare the total cross sections
obtained with the popular Donnachie--Landshoff fit \cite{Donnachie93}.
In Fig. \ref{xsd2126981}  we plot the fitted single diffractive
cross sections  together with data 
 \cite{Albrow76,Armitage82,Amos93a,Abe94c,Bernard87b}. 

\begin{table}[hbt]
\caption{
     DTUJET99 model parameters obtained with an energy dependent
$p_{\perp thr}$ .}
\label{DTUJET97fit}
\begin{center}
\begin{tabular}{|l||c|c|c|c|c|c|} \hline
PDF set& $g^2 (mb)$ & $\alpha (0)$ &$\alpha^{\prime}~[GeV^{-2}]$
&$b~[GeV^{-2}]$&$b_h~[GeV^{-2}]$
& $\lambda$  \\ \hline \hline
GRV--LO--98
& 55.16  & 1.0513 & 0.325 & 1.114 & 1.70 & 0.585  \\ \hline
\end{tabular}
\end{center}
\end{table}
%
 
 The single diffractive  cross section becomes energy independent at
 high energies in
 DTUJET and DPMJET.
 This follows from  the triple Pomeron cross section chosen in
 \cite{DTUJETPR92} 
 without the energy rise according to the supercritical Pomeron.
 (As explained in detail elsewhere DPMJET follows in
 the unitarization procedure DTUJET.) 
  Such an approximation is not necessary in
 the Dual Parton Model. It has been shown by Engel within
the PHOJET Dual Parton Model \cite{Engel95a} that the proper inclusion
of the supercritical triple Pomeron cross section as well as 
higher order terms like the double pomeron graph into the unitarization
procedure also results in a single diffractive cross section becoming
rather energy independent at high energy in agreement to the data.
Similar results were obtained by Gotsman, Levin and Maor
\cite{Gotsman93}. On a phenomenological basis this saturation of the
single diffractive cross section was described by Goulianos 
\cite{Goulianos95} using the term of a {\it renormalized flux}.

\begin{figure}[thb]
\begin{center}
 \psfig{figure=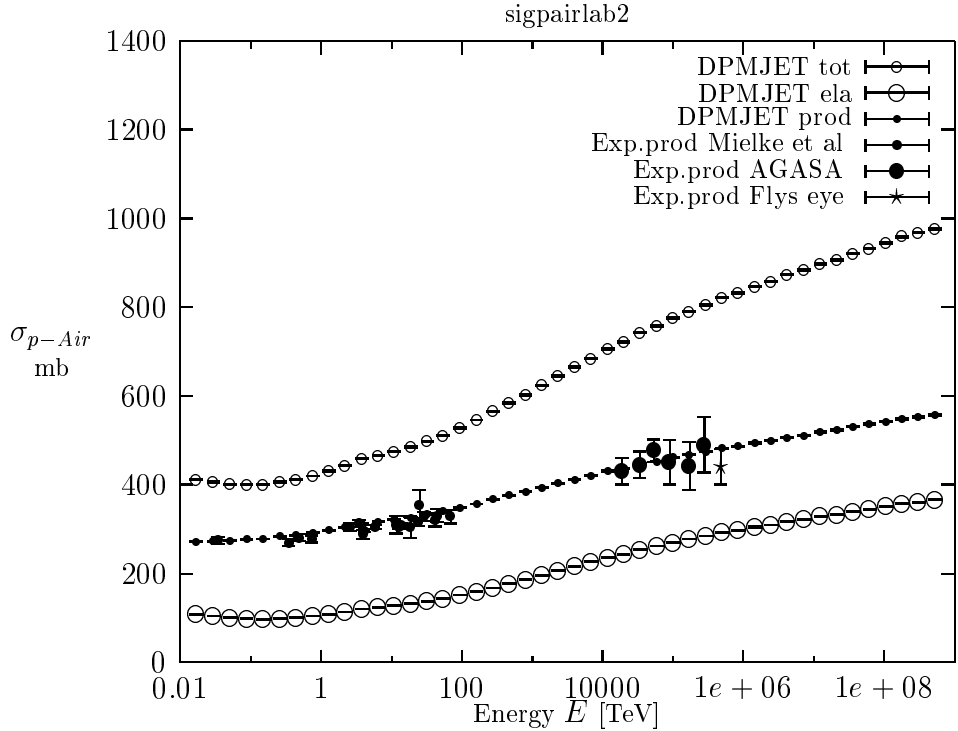}
\end{center}
\vspace*{-3mm}
\caption{The inelastic (production) 
cross section \protect$\sigma _{p-Air}$ calculated
by DPMJET--II.5 as
function of the laboratory collision energy (from 0.01 TeV up to
10\protect${}^9$ TeV) compared to
experimental data collected by Mielke et al. \protect\cite{Mielke} and
to experimental data from the AGASA \protect\cite{Honda93} and Fly's eye
\protect\cite{Baltrusaitis84} experiments as presented
in corrected form in a recent paper by Block, Halzen and Stanev
\protect\cite{Block99}.
Given are also the calculated
total and elastic cross sections.
\protect\label{sigpair2}
}
\end{figure}
\begin{figure}[thb]
\begin{center}
 \psfig{figure=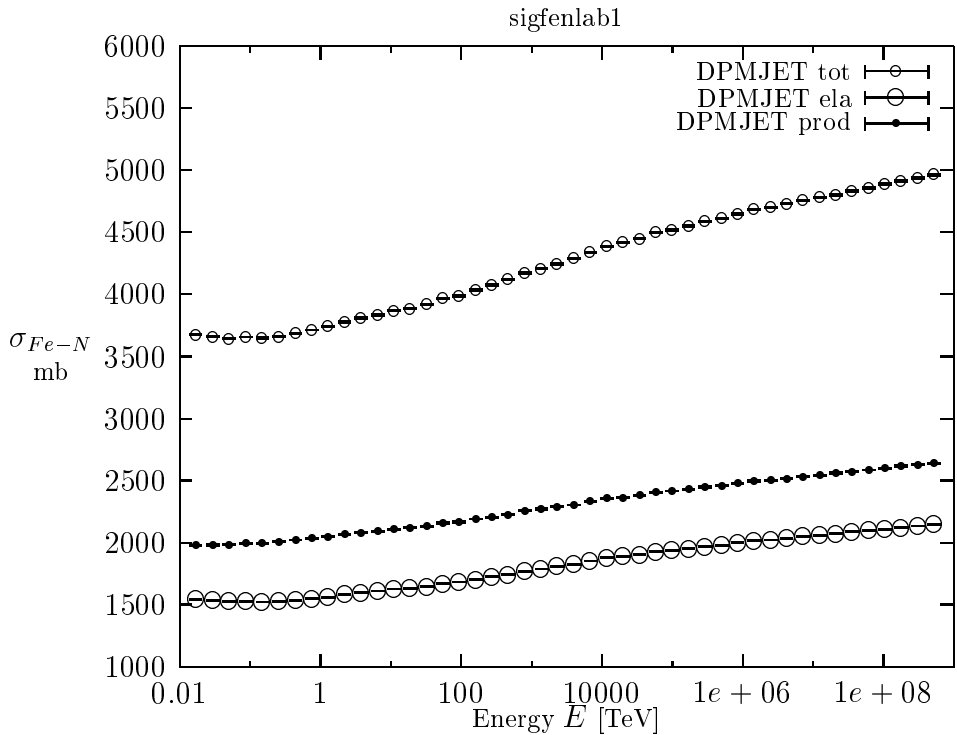}
\end{center}
\vspace*{-3mm}
\caption{The inelastic (production) 
cross section $\sigma _{Fe-N}$ calculated
by DPMJET--II.5 as
function of the laboratory collision energy (from 0.01 TeV up to
1.E9 TeV). 
Given are also the calculated
total and elastic cross sections.
\protect\label{sigfeairl}
}
\end{figure}

%
\subsection{The Dual Parton Model for hadron production in
hadron--nucleus and nucleus--nucleus collisions}

In the following we discuss new features in DPMJET--II.5,
for a more detailed description of the model as applied in the code
we refer to 
Refs.~\cite{RaRiHA,RaRiHA1,Ranft88a,Moehring91,KMR92,MRCT93,CTR1,Ranft94b,Ferrari95a,Ferrari96a,Battistoni96a}.
%
%
\subsubsection{The Monte Carlo realization of the
	 dual parton model DPMJET
	 for hadron-nucleus and nucleus-nucleus collisions}

The model starts from the impulse approximation
for the interacting nuclei -- i.e. with a frozen discrete spatial
distribution of nucleons sampled from standard density distributions
\cite{Shma88,Barrett77}.
	The primary interaction of the incident high--energy projectile
	proceeds via  $n$ elementary collisions between
        $ n_p = n_A $  and  $ n_t = n_B $  nucleons from the projectile
        (for incident hadrons $n_p=1$~) and the target nuclei, resp, .
        Actual numbers $n, n_p$  and  $n_t$  are sampled on the basis
        of Glauber's multiple scattering formalism
        using the Monte Carlo algorithm of Ref.\cite{Shma88}.

The Glauber model, which is part of DPMJET--II allows to
calculate the inelastic hadron--nucleus cross sections. The needed
ingredients of  
 this calculation are the nuclear geometry and the
elementary hadron--nucleon scattering amplitude. 

The nuclear matter distribution is parametrized as follows:

For most nuclei (untill DPMJET--II.3/4 all nuclei) we parametrize the
average nuclear radius (in fm)
\begin{equation}
\label{nucrad}
c=\sqrt{r^2} = 1.12 N_a^{0.33}
\end{equation}
This parametrization is however not reliable for some light nuclei
important for Cosmic Ray collisions. Therefore we replace
(\ref{nucrad}) for these nuclei by the values given in Table 2
 \cite{Barrett77}.

{\bf Table 2.}
Average nuclear radii used in DPMJET--II.5 \cite{Barrett77} (in fm).
\vskip 5mm
\begintable
%
  $N_a$   | 9| 10 |11 |12 |13 |14 |15 |16 |17 |18 ~\cr
   $\sqrt{r^2}$ | 2.52| 2.45 |2.37 |2.45 |2.44 |2.55 |2.58 |2.71 |2.66 |2.72 \endtable
\vskip 10mm 
The nuclear density distribution is parametrized for most nuclei 
as the Fermi
distribution
\begin{equation}
\label{fermidist}
F(r)=\frac{1}{1+\exp{\frac{r-c}{0.545}}}
\end{equation}
However, for $N_a$ between 9 and 18 we use again more precise density
distributions as given in Ref.\cite{Barrett77}.

The
elementary hadron--nucleus scattering amplitude is parametrized as
follows:

Energy dependent quantities enter the Glauber approach via the profile
function of elastic hadron-nucleon scattering,
\begin{equation}
    \gamma_{hN}(b) = {1 \over 2 \pi i p}
                \int d^2q \exp(i \vec{q} \cdot \vec{b})
f_{hN}(\vec{q}),
\end{equation}
i.e. the amplitude of elastic hadron-nucleon scattering in the impact
parameter representation (with $\vec{q}$ denoting the lateral, i.e.
two-dimensional momentum transfer).
In their Monte Carlo realization of Glauber's approach Shmakov et al.
\cite{Shma88}
apply the high-energy approximation of the profile
function,
\begin{equation}
    \gamma_{hN}(b) = {\sigma^{tot}_{hN} \over 4 \pi a}
                     \Biggl( 1 - i {\Re f_{hN}(0) \over \Im
f_{hN}(0)}
                          \Biggr) e^{-{b^2 \over 2a}},
\end{equation}
with parameters $\sigma^{tot}, a$ and
 $\rho=\Re f_{hN}(0)~/~\Im f_{hN}(0)$  appropriate for the description
of nucleus-nucleus interactions at energies of several
GeV per nucleon.
(This parametrization  
 corresponds to a differential
cross section  $d\sigma/dt \simeq \sigma_{tot} \exp(a\cdot t)$ with
$ t \simeq -\vec{q}^2$.)
 
However, the energy dependence of the elastic hadron-nucleon scattering
amplitudes will influence the properties of hadron--nucleus and
nucleus--nucleus
scattering. In particular, the number of individual high-energy
hadron-nucleon interactions ($n, n_A, n_B$) will increase with rising
energy, hence the multiplicity will increase stronger than to be
expected from the energy dependence of single hadron-hadron
interactions.
 
Guided by the data collected in Ref. \cite{Castaldi},
we apply the following parametrizations for
the slope parameter $a$ :
$   a  = 8.5 \ (1 + 0.065 \ln s)$ for nucleon-nucleon
collisions above $s$ = 50 GeV${}^2$,
$   a  = 6.2 \ (1 + 0.13 \ln s)$ for nucleon-nucleon
collisions below $s$ = 50 GeV${}^2$ 
and
$  a =    6.0 \ (1 + 0.065 \ln s)$ for $\pi$-- and K--nucleon
collisions. 
We use for
 the ratio $\rho$ of the real and imaginary
part of the elastic scattering amplitude:
 $ \rho  =
                         -0.63 + 0.175 \ln\sqrt s $ for the energy
region
$   3.0 \leq \sqrt s  \leq 50. $ and $\rho=  0.1 $ in the energy region
$ \sqrt s \geq 50 $ GeV in
 nucleon-nucleon scattering and $\rho =
                   0.01 $ for $\pi$-- and K--nucleon scattering.

The energy dependence of the total cross sections is described by the
fits of the Particle Data Group \cite{revpartprop};
 at energies beyond
the range of the actual parametrization of the $pp$ cross section the
one for $\bar{p}p$ is applied and at energies even higher  we use
the total cross--sections as calculated by the two--component 
Dual Parton Model
for hadron--hadron collisions, see also Fig.\ref{xs2126981}.

The same information is also needed to construct the inelastic
events and indeed  for nuclear collisions   
each run of DPMJET starts with a calculation
of the inelastic cross section. 

  A new option (XSECNUC) has been added to DPMJET--II.5 to
  calculate a table of hadron--nucleus or nucleus--nucleus
  cross sections as function of the collision energy. The
  routine XSGLAU (this is  a routine adopted from DTUNUC--II
  \cite{Roesler96b,Roesler99}) 
  using the method according to \cite{Shma88}) 
  calculates the following cross sections:
 \vskip 5mm

  (1)$\sigma_{tot}^{A-B}$
 \vskip 5mm

  (2)$\sigma_{el}^{A-B}$~~~~~~~~~~~~~~~~ $A+B\rightarrow A+B$
 \vskip 5mm

  (3)$\sigma_{quasi-el-3}^{A-B}$ ~~~~~~~~~~$A+B \rightarrow A+X$, excluding
  (2)
 \vskip 5mm

  (4)$\sigma_{quasi-el-4}^{A-B}$ ~~~~~~~~~~$A+B \rightarrow X+B$, excluding
  (2)
 \vskip 5mm

  (5)$\sigma_{quasi-el-5}^{A-B}$ ~~~~~~~~~~~$A+B \rightarrow X$, excluding
  (2),(3) and (4)
 \vskip 5mm

  (6)$\sigma_{prod}^{A-B}$=~~$\sigma_{inel}^{A-B}$=~~$\sigma_{tot}^{A-B}$--~~ 
  $\sigma_{el}^{A-B}$--~~ $\sigma_{quasi-el-3}^{A-B}$--~~
  $\sigma_{quasi-el-4}^{A-B}$--~~ $\sigma_{quasi-el-5}^{A-B}$ 
 \vskip 5mm

In Fig.\ref{sigpair2} and \ref{sigfeairl} 
we present  the cross sections 
(only $\sigma_{tot}^{A-B}$, $\sigma_{el}^{A-B}$ 
and $\sigma_{prod}^{A-B}$ )
as calculated by DPMJET--II.5 using
the XSECNUC option for
p--Air  and Fe--N which are relevant for calculating
the Cosmic Ray cascade as function of the lab energy from 0.02
TeV to 10${}^9$ TeV.  The quasi-- elastic cross sections 
$\sigma_{quasi-el-i}^{A-B}$ are only contained in these plots only in 
in $\sigma_{tot}^{A-B}$.
The inelastic DPMJET p--Air cross sections ~$\sigma_{inel}^{A-B}$ 
are compared to the
experimental data collected by Mielke et al. \cite{Mielke} and to
experimental data from the AGASA \cite{Honda93} and Fly's eye
\cite{Baltrusaitis84} experiments as presented
in corrected form in a recent paper by Block, Halzen and Stanev
\cite{Block99}.

\subsection{Implementation of new DPM diagrams for an improved
description of baryon stopping in hadron--nucleus and nucleus--nucleus
collisions}

\subsubsection{Diquark fragmentation, the popcorn mechanism}

The fragmentation of diquarks is slightly more complicated than
the fragmentation of a quark jet.
As justified by  Rossi and Veneziano \cite{Rossi97} the baryon can
be pictured as made out of three quarks bound together by three
strings which join in a so called string junction point.
In diagrams we can characterize the baryons 

(i)by the three quarks
and the string junction or 

(ii)by a quark and a diquark, in this case the string junction always
goes with the diquark.

In all the diagrams discussed in
this section we will either plot the quark and the diquark or, if the
diquark breaks, the three quarks and the string junction. The quarks and
the two quarks of a diquark are plotted as solid lines, the string
junction is plotted as a dashed line.

\begin{figure}[thb] \centering
\begin{turn}{-90}
 \psfig{figure=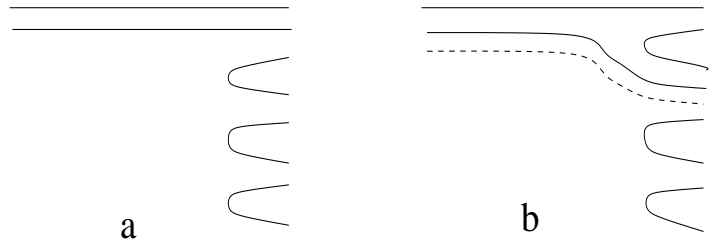,width=4.0cm,height=10.0cm}
\end{turn}
 \vspace*{2mm}
 \caption{{\bf (a) }Conventional diquark fragmentation, the baryon is
 produced in the first fragmentation step. {\bf (b) }The popcorn
 mechanism: a meson is produced in the first fragmentation step, the
 baryon appears in the second  or later fragmentation
 steps. 
 We plot the lines for the two quarks of the diquark as solid lines,
  the string junction is plotted as a dashed line.
 \label{popc}
 }
 \end{figure}

There are 
 two possibilities for the first fragmentation step of a diquark, see
 Fig.\ref{popc}.
 Either we
get in the first step a baryon, which contains
both quarks of the diquark and the string junction or we get  in the
first step a meson containing only one of the two quarks and the
baryon is produced in one of  
the following fragmentation step. This
mechanism is well known, it is presented in the review on the
Dual Parton Model
\cite{capdpm} and it was investigated 
for instance in \cite{SUKHA,KOPEL}.
 This mechanism was implemented
from the beginning in the 
BAMJET--fragmentation code \cite{BAMJET1,BAMJET}
used previously in DPMJET. 
This mechanism is also implemented under the name {\it popcorn}
fragmentation in the Lund chain fragmentation model JETSET
\cite{JETSET,AND85}. 
\\
 What happens in the model with the popcorn mechanism compared to the
model without can be most easily seen looking at the proton
rapidity distribution in p--p collisions.
The two maxima in the target and projectile
fragmentation region of the proton rapidity distribution shift by
about half a unity to the center,  these maxima become
wider and  correspondingly the dip in the center is reduced.
At the same time the Feynman $x$ distributions of mesons get
a component at larger Feynman $x$. The effects in hadron--nucleus and
nucleus--nucleus collisions are quite similar.
The popcorn mechanism is however not enough to explain the baryon
stopping observed experimentally in hadron--nucleus and nucleus--nucleus
collisions
 \cite{NA35FIN,Alber98}, this will be discussed in more detail  in
 Section III.

\subsubsection{New diquark breaking DPM diagrams 
in hadron--nucleus and nucleus--nucleus
collisions}

\begin{figure}[thb] \centering
\begin{turn}{-90}
 \psfig{figure=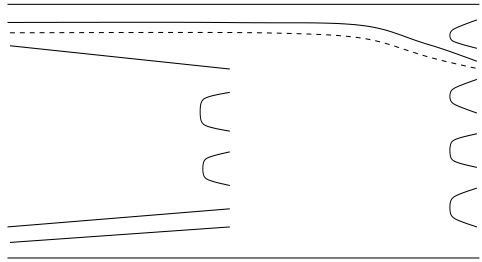,width=4.0cm,height=7.0cm}
\end{turn}
 \vspace*{2mm}
 \caption{The first C--K diquark breaking mechanism \protect\cite{Capella96}
 plotted for a nucleon--nucleon collision.
 \protect\label{diqb}
 }
 \end{figure}

 New diquark breaking DPM--diagrams mainly of interest in
 hadron--nucleus and nucleus--nucleus collisions were 
 proposed by Capella and Kopeliovich
 \cite{Capella96} and investigated in detail by Capella and
 Collaborators \cite{Capella99a,Capella99,Capella99b}. 
 Similar ideas were discussed by Vance and
 Gyulassy\cite{Vance99}.
Capella and Kopeliovich\cite{Capella96} did discuss 
in detail their first diquark
breaking mechanism (see Fig.\ref{diqb}, where this mechanism is
characterized for nucleon--nucleon collisions), in this case the
valence--diquark breaks into the two quarks, the baryon is produced in
the second or in later fragmentation steps. 

Looking only at  the diagrams, 
one does not see any difference between the popcorn
 mechanism and the first  Capella--Kopeliovich (C--K) diquark breaking
 mechanism.
 However, the mechanism of this first  
 (C--K) diquark breaking
 mechanism  differ in detail from the popcorn mechanism discussed
 above. Nevertheless, implementing the first 
 C--K mechanism in DPMJET did not give
 any new feature of baryon stopping, which could not also 
 be obtained from
 the popcorn mechanism. Therefore, 
 we continue to use in DPMJET--II.5 the
 popcorn mechanism instead of the C--K mechanism, there is no need to
 use both effects.

\begin{figure}[thb] \centering
\begin{turn}{-90}
 \psfig{figure=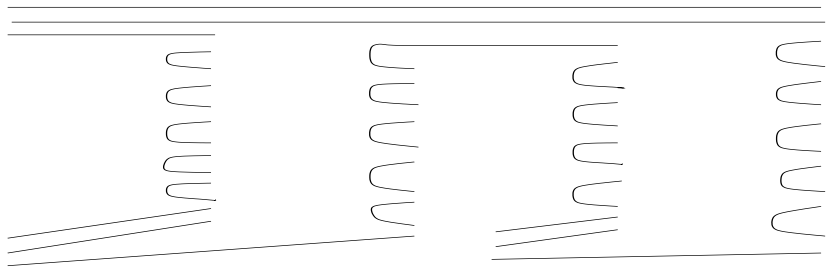,width=4.0cm,height=12.0cm}
\end{turn}
 \vspace*{2mm}
 \caption{The diquark--conserving diagram for a nucleon--nucleus
 collision with two participants of the target nucleus.
 \protect\label{diqc}
 }
 \end{figure}

 More interesting is the second C--K mechanism, which was proposed but 
 not discussed in detail in \cite{Capella96}. This mechanism was 
 discussed in detail by Capella and 
 Collaborators\cite{Capella99a,Capella99,Capella99b}. In
 Fig.\ref{diqc} we plot first the diquark--conserving diagram for a
 nucleon--nucleus collisions with two participants of the target
 nucleus. This is the traditional way for such a collision in the DPM.

\begin{figure}[thb] \centering
\begin{turn}{-90}
 \psfig{figure=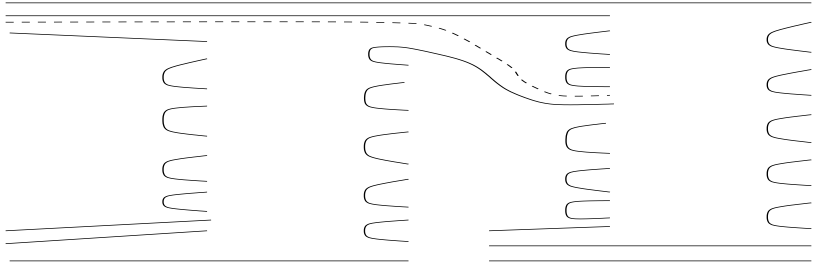,width=4.0cm,height=12.0cm}
\end{turn}
 \vspace*{2mm}
 \caption{The Glauber sea quark mechanism of baryon stopping GSQBS for a
 nucleon--nucleus collision with two 
 participants of the target nucleus. This
 is the second C--K diquark--breaking mechanism \protect\cite{Capella96}.
 \protect\label{diqsea}
 }
 \end{figure}

In Fig.\ref{diqsea} we plot the second C--K diquark--breaking diagram
for the same collision. 
Now the second 
valence quark from the broken diquark 
in the first C--K diagram 
in Fig.\ref{diqb} is
replaced by a Glauber 
sea quark from the nucleon projectile. Therefore, we will
call the mechanism the Glauber 
sea quark mechanism of baryon stopping GSQBS. The probability of such a
diquark splitting rises if the considered nucleon is involved in more
than two interactions. The GSQBS has been
implemented in DPMJET--II.5 and we will see in Section III, that in
nucleon--nucleus collisions and nucleus--nucleus collisions we are able
with this mechanism 
to fill the dip in the baryon rapidity distributions at central rapidity
in agreement to the experimental data. As discussed already in detail in
\cite{Capella99,Capella99b} 
this mechanism also contributes to increase the Hyperon
production in nucleon--nucleus and nucleus--nucleus collisions.

\subsubsection{The Casado diagram 
}

\begin{figure}[thb] \centering
\begin{turn}{-90}
 \psfig{figure=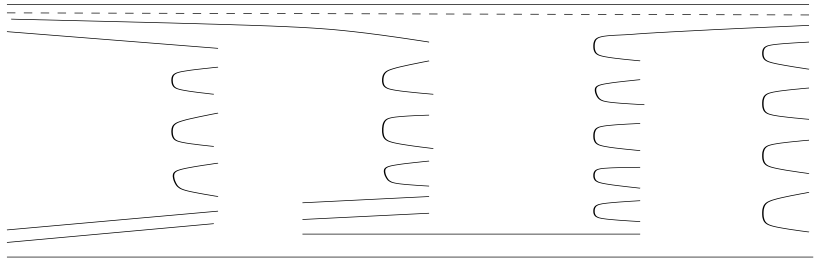,width=4.0cm,height=12.0cm}
\end{turn}
 \vspace*{2mm}
 \caption{The diagram introduced by Casado \protect\cite{Casado99}
 for a nucleon--nucleus
 collision with two participants of the target nucleus. The diquark
 contains one Glauber sea quark.
 \protect\label{casado}
 }
 \end{figure}

A  new diagram was also introduced by Casado \cite{Casado99}. The
diagram is plotted again for a nucleon--nucleus collision with two
participants of the target nucleus
in Fig. \ref{casado}. Here the fragmenting diquark contains a Glauber
sea quark and hadronizes like a valence diquark producing baryons mainly
in the fragmentation region. However, the flavor content of the baryon 
is changed. This diagram gives another contribution to hyperon
production in nuclear collisions. We have implemented the Casado diagram
in DPMJET--II.5 and it is used in addition to the GSQBS diagram for all
nuclear collisions compared in  Section  III to data.

\subsubsection{
Two extrapolations of DPMJET--II.5 to Cosmic Ray energies.
}

\begin{figure}[thb] \centering
\begin{turn}{-90}
 \psfig{figure=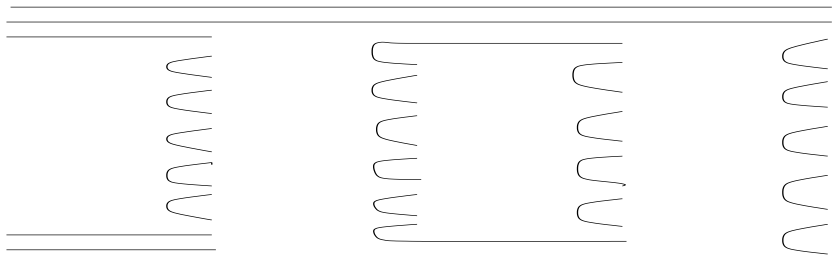,width=4.0cm,height=12.0cm}
\end{turn}
 \vspace*{2mm}
 \caption{Standard DPM diagram for a nucleon--nucleon interaction with
 one additional soft secondary interaction 
 induced by the unitarzation procedure.
 There is one valence--valence and one sea--sea interaction, each
 represented by a pair of chains.
 \protect\label{normal2}
 }
 \end{figure}

\begin{figure}[thb] \centering
\begin{turn}{-90}
 \psfig{figure=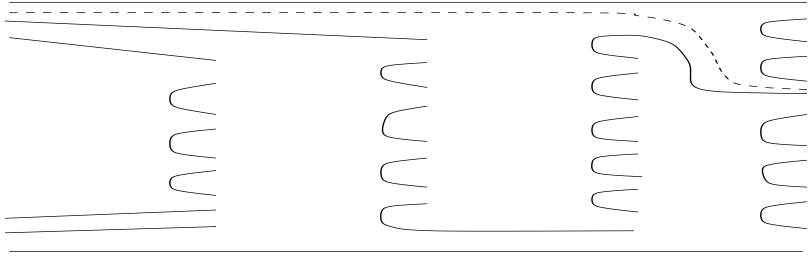,width=4.0cm,height=12.0cm}
\end{turn}
 \vspace*{2mm}
 \caption{New DPM diagram for a nucleon--nucleon interaction with one
 additional soft secondary interaction 
 induced by the unitarzation procedure. The
 diquark is split and an unitarity sea quark is used similar to Fig.
 \protect\ref{diqsea} to shift the baryon in one of the chains. We call this the
 unitary sea quark mechanism for baryon stopping USQBS.
 \protect\label{new2}
 }
 \end{figure}

At high energies we have multiple collisions even in hadron--hadron
collisions due to the unitarization procedure. We call the sea quarks at
the ends of the additional chains in 
this case {\it unitary sea quarks}. The
Glauber sea quarks are needed in nuclear collisions already at rather
low energies, for instance at the energies of heavy ion collisions at
the CERN--SPS. In contrast to this, unitary sea quarks appear in
significant numbers in
hadron--hadron and nuclear collisions only at rather high energies, for
instance at the energies of the CERN--SPS collider or the TEVATRON
collider, they are important in the Cosmic Ray energy region.

With the unitary sea
quarks at the ends of the chains from the secondary collisions we obtain
a new mechanism for baryon stopping, which will become effective at very
high energies.

In Fig.\ref{normal2} we plot the standard DPM diagram for a
nucleon--nucleon interaction with two soft  interactions induced
by the unitarization procedure. There is one valence--valence and one
sea--sea interaction, each
represented by a pair of chains. In analogy to Fig. \ref{diqsea} we
construct from this the new diagram for baryon stopping in Fig.
\ref{new2}.
  The
 diquark is split and an unitary sea quark is used 
 to have the  baryon only in the second or later fragmentation steps 
 in one of the chains. We call this the
 unitary sea quark mechanism for baryon stopping USQBS. Also here the
 probability for such a diquark splitting rises if there are more than 2
 interactions of the hadrons involved. Obviously, 
 this mechanism leads to  Feynman $x$ distributions of baryons in
 p--p collisions becoming  softer and Feynman $x$ distributions of
 mesons becoming  harder than without the USQBS mechanism.

 In contrast to the  GSQBS mechanism, which leads already
 to effects in nuclear collisions 
 at the energy of the CERN--SPS, we have at present
 no data available to prove that this USQBS mechanism is a valid
 extension of the DPM. There is no data for baryon stopping in
 proton--proton or antiproton--proton collisions at collider energies.
 Unfortunately, the fragmentation region at large Feynman $x$ has not
 been investigated experimentally with enough detail 
 at any of the hadron--hadron
 colliders.
However, if the GSQBS mechanism is the correct mechanism responsable for
the baryon stopping effects found in nuclear collisions, then also the
USQBS mechanism should modify the collisions at collider energies and
beyond. In  Section IV, where we present the properties of DPMJET--II.5
at Cosmic Ray energies we present always two extrapolations of the
model. In one version (shortly characterized as version 55 in the plots)
we have both the GSQBS and USQBS mechanism, in a second version
(characterized in the plots as
version 50) we have only the GSQBS mechanism. Only collider experiments
on baryon stopping or Cosmic Ray observations can prove, which version
is the better one. From theoretical prejudices however there would be
the claim, that the version with both the GSQBS and USQBS mechanisms
is to be prefered.

\subsubsection{New parameters connected with the diquark breaking
diagrams
}

For each of the new diquark breaking diagrams described in this Section
we have to introduce a new parameter. These parameters give the
probability for the diquark breaking mechanisms to occur, 
given a suitable sea
quark is available and given that the diquark breaking mechanism is
kinematically allowed. For an original diquark--quark chain of small
invariant mass, which originally 
just fragments into two hadrons, the diquark
breaking is often not allowed at small energies.

The optimum values of the new parameters are determined by comparing
DPMJET--II.5 with experimental data on net--baryon distibutions. We get
roughly for all of these probabilities values around 0.5. The
definitions of these parameters and their default values in DPMJET--II.5
are given in the companion paper \cite{Ranft99b}.

\subsection{Realization of the intranuclear cascade process}

Nothing has been changed in DPMJET--II.5 against DPMJET--II.3/4 with
respect to the formation zone intranuclear cascade, nuclear
evaporation and fragmentation,  $\gamma$ deexcitation of the
residual nuclei and the Cronin effect. All of this was described
in detail in
\cite{dpmjet2324,Ferrari95a,Ferrari96a} .

\subsection{Production of strange particles}

Studies of strangeness production within
this model were  given in \cite{MRCT93,CTR1,CTR94}.
Enhanced generation of strange particles,
in particular of
strange antibaryons, has been proposed
as a signal for the formation of quark gluon plasma in dense
hadronic matter~\cite{rafel,rafel1}.
Recent data from experiments at the CERN SPS have already
been interpreted
within this scheme.
However, we find it worthwhile to pursue the study of
tested conventional models
without QGP formation like the DPM before drawing final
conclusions.
The DPM  is an independent string model.
 Since the individual strings are universal building
blocks of the model, the ratio of {\it produced} strange particles
over non--strange ones will be approximately the same in all
 reactions. However, since some strings contain sea quarks at one
or both ends and since strange quarks are present in the proton sea,
it is clear, that, by increasing the number of those strings, the
ratio of strange over non--strange particles will increase.
This will be the case for instance, when increasing the
centrality in a nucleus--nucleus collision. It is obvious, that
the numerical importance of the effect will depend on the assumed
fraction of strange over non--strange quarks in the proton sea.
 The rather extreme case leading to a maximum increase of strangeness
would be to assume a SU(3) 
symmetric sea (equal numbers of $u$, $d$ and $s$
flavors).
 We express the
amount of SU(3) symmetry of the sea chain ends by our parameter
$s^{sea}$ defined as
$s^{sea} =  {2<s_s>}/({<u_s>+<d_s>})$
where the $<q_s>$ give the average numbers of sea quarks at the
sea chain ends. Usually, DPMJET uses $s^{sea}$ = 0.5.
\\
There are a number of effects, which change the number of strange
hadrons, especially strange and multistrange baryons 
and antibaryons in the model: 

(i)The presence of sea $qq$ and $\bar q \bar q$ diquarks at sea--chain
ends \cite{Ranft93a,Moehring93a}.

(ii)The popcorn effect.

(iii)The diquark breaking diagrams discussed in Section II.C, see also
 \cite{Capella99a,Capella99,Capella99b}. 

(iv)The secondary interactions of co--moving produced particles
 \cite{Capella95,Capella95b}, see also Section II.G.

(v)Effects of string fusion \cite{Merino,Moehring93b}
and percolation of strings \cite{Armesto96,Braun97}.

It is certainly beyond the scope of the present paper, to discuss all of
these effects in detail. The effects (i) to (iv) are contained in
DPMJET--II.5.

\subsection{Single diffractive cross sections in hadron--nucleus and
nucleus--nucleus collisions }

Let us first summarize the sampling of single 
diffractive events in in DPMJET
in hadron--hadron collisions.

Single diffraction within the Dual Parton Model 
was studied in detail and compared to
experimental data in \cite{RER93,Ranft94b}.
 Single diffraction dissociation
is represented by a triple--Pomeron
exchange (high mass single diffraction) and a low mass component
(low mass single diffraction) \cite{DTUJETPR92}.

Diffractive processes characterized by the excitation
of an initial hadron to intermediate resonances (low mass
diffractive interactions) are introduced via a two channel
eikonal formalism.
\\
High mass single diffractive events are sampled cutting the triple--Pomeron
graph . The excited system consists of two
chains stretched between the constituents of the excited hadron and a
$\rm{q\bar{q}}$--pair of the Pomeron emitted from the 
hadron at the upper vertex.

Corresponding to $M_{D}^{2}=x_{D}s$ and to the known
$1/M_{D}^{2}$--behaviour
of the diffractive differential 
cross section $d^{2}\sigma_{sd}/dtdM_{D}^{2}$
we sample the sum of the momentum fractions of the $\rm{q\bar{q}}$--pair
$x_{D}=x_{q}+x_{\bar{q}}$ from a $1/x_{D}$--distribution.
Assuming that inelastic diffraction dissociation puts virtual hadronic
states on the mass shell we limit this selection by the coherence 
condition.
\begin{equation}
M_{D}^{2}/s\le (m_{p}R)^{-1} \qquad\mbox{with}\qquad R \simeq
m_{\pi}^{-1}
\end{equation}
where $\sqrt{s}$ is the c.m. energy of the collision and $R$ is the
interaction radius.
\\
An excited low mass single diffractive event is represented by a chain
connected to the valence partons of one hadron. 
The invariant mass $M_{D}$
is fixed in the same way as above.

 We turn to single diffractive cross sections in hadron--nucleus and
 nucleus--nucleus collisions.

 In DPMJET the simple approximation is used: single diffraction in
 nuclear collisions occurs only in Glauber collisions with only one
 participant of projectile as well as target ($N_A$ = $N_B$ = $N$ = 1).
 In such collisions also the total number $N$ of Glauber collisions is
 equal to one. In the first step of 
 the $N_A$  = $N_B$ = $N$ = 1 collision we have just a
 single hadron--hadron collision, single diffraction  occurs, if
 this single collision is a diffractive collision.  In the second step
 of the single diffractive hadron--nucleus or nucleus--nucleus collision
 we have the formation zone cascade  in the participating nuclei
 followed by nuclear evaporation. There might be single diffractive
 collisions also in Glauber collisions with $N$ larger than one. In such
 cases it is rather unlikely, that all collisions will be diffractive.
 If we have one single diffractive collision together with one or
 several nondiffractive collisions, the event just looks like a
 nondiffractive one. We neglect in DPMJET the small fraction of
 diffractive events, where all of these Glauber collisions are
 diffractive.

 DPMJET allows in hadron--nucleus and nucleus--nucleus collisions like
 in hadron--hadron collisions to chose all collisions or only
 nondiffractive or only diffractive collisions. DPMJET always calculates
 the fraction of diffractive collisions to the inelastic cross section.

\begin{figure}[thb]
\begin{center}
 \psfig{figure=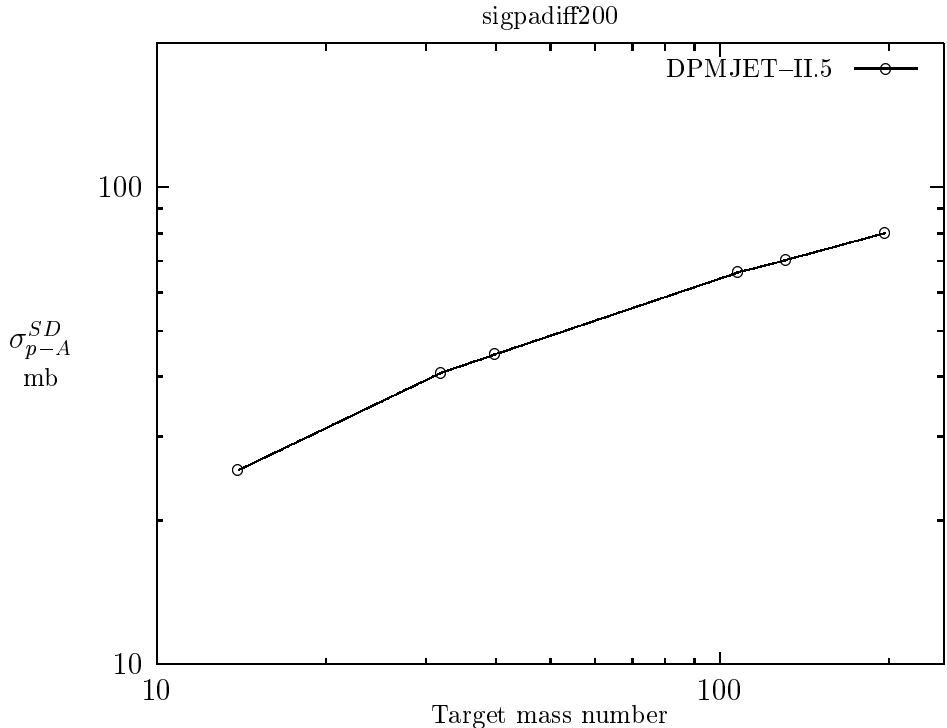}
\end{center}
\vspace*{-3mm}
\caption{Single diffractive cross sections calculated with DPMJET--II.5
in proton--nucleus collisions.
  The single diffractive cross sections given are  for
 diffraction on the target as well as the projectile side. 
\protect\label{sigpadiff200}
}
\end{figure}

 In Fig.\ref{sigpadiff200} we plot the single diffractive cross sections
 in proton--nucleus collisions as function of the nuclear mass number
 $A$. The single diffractive cross sections given are always for
 diffraction on the target as well as the projectile side. 
 As to be expected we find $\sigma^{h-A}_{SD} \approx A^{1/3}$.
 Single diffractive nucleus--nucleus cross sections $\sigma^{A-B}_{SD}$
 are found to be rather similar to the h--Max(A,B) diffractive cross
 sections. In Table 3  
 we give only some examples of single diffractive nucleus--nucleus cross
 sections obtained from DPMJET--II.5.
 
{\bf Table 3.}
Diffractive cross sections in nucleus--nucleus collisions (in
mb).
  The single diffractive cross sections given are  for
 diffraction on the target as well as the projectile side. 
\vskip 5mm
\begintable
 Reaction   | $\sigma^{A-B}_{SD}$  ~\cr
  S--S         | 50.5  \cr
  S--Ag        | 60.9  \cr
  S--Au        | 67.6  \cr
  Pb--Pb        | 72.8  \endtable
 \vskip 10mm 


\subsection{Final state interactions of co--moving secondaries in
nuclear collisions}

Analyzing the rapidity distributions of produced $\Lambda$ and $\bar
\Lambda$ perticles in central heavy ion collisions at CERN--SPS energies
within the Dual Parton Model Capella et al \cite{Capella95,Capella95b}
noted the need for secondary interactions of co--moving secondaries to
understand the data. They only introduced the following secondary
interactions
\begin{equation}
\pi + N \rightarrow K + \Lambda
\end{equation}
and
\begin{equation}
\pi + \bar N \rightarrow K + \bar \Lambda
\end{equation}
A reasonable cross section for these reactions 
of $\sigma \approx$ 1.5 mb was needed in \cite{Capella95,Capella95b}
to understand the data. 
 All of
this was done in  \cite{Capella95,Capella95b} using analytical methods,
not a Monte Carlo event generator. 
Final state interactions of co--moving secondaries was also used in
\cite{Armesto97,Armesto99} to explain the $J/\psi$ supression in Pb--Pb
collisions.

The method of Capella et al  \cite{Capella95,Capella95b} was implemented
in DPMJET already in 1995. In DPMJET the method is used to modify the
Monte Carlo events 
and we use only a cross section of
$\sigma \approx$ 1.0 mb. 
First results reproducing essentially the
results of  \cite{Capella95,Capella95b} are described shortly in
\cite{Ranftsare95} and in more detail in a unpublished code write--up
\cite{Ranft95b}.  We would like to stress, this method as implemented 
 with only the two reactions given above, can only be
considered as a first preliminary step. A better method of final state
interactions should be implemented as a Monte Carlo method from the
beginning and  more types of 
secondary interactions should be taken into account
including the reverse reactions.
With the present method we can not expect to obtain reasonable results
in situations with much higher secondary particle densities than in the
heavy ion collisions at the CERN--SPS energy. Therefore, it is not
recommended to use the secondary interaction option of DPMJET at RHIC
or CERN--LHC energies.

\subsection{Central nucleus--nucleus collisions in DPMJET}

The sampling of a Glauber event according to the method of \cite{Shma88}
proceeds as follows: First the impact parameter $B$ of the hadron--nucleus
or nucleus--nucleus collision is sampled from a distribution determined
at the beginning by a Monte Carlo method. In a second step the  number
of interacting projectile and target nucleons $N_A=N_p$ and $N_B=N_t$
and the total number of Glauber collisions $N$ are sampled.

We might define in DPMJET a central collision by different methods:

(i)Allowing only impact parameters $B \leq B_{thr}$.

(ii)Demanding, that a certain minimum fraction of the projectile or
target nucleons takes part in the collision  for instance 
$N_A=N_p \geq \alpha A$,
where $\alpha$ is a number not far from unity.  

Experiments  investigating central nucleus--nucleus collisions   are
usually not able to determine reliably the impact parameters of the
collisions. They might be able to determine the number of interacting
projectile nucleons, for instance using a zero--degree calorimeter. 
Most often one finds however the central collision characterized 
by giving which fraction $F_{central}$ of all events 
are contained in the sample of most 
central collisions considered.

In order to keep contact to the $F_{central}$ definition 
we proceed in DPMJET 
 in an iterative way.  DPMJET is run  rejecting all events
with  $N_A=N_p \leq \alpha A$ for a given $\alpha$. 
At the same time we keep track which fraction of the 
events $f_{\alpha}$ is
accepted in the run. $\alpha$ is iterated untill $f_{\alpha} \approx
F_{central}$.

\subsection{Restrictions of DPMJET--II}

  DPMJET--II.5   works for hadron--hadron and 
  hadron--nucleus collisions above
50~MeV kinetic energy, it has been compared to data on average
multiplicities down to about 1 GeV and to data on Feynman $x$
distibutions down to 3-4 GeV see \cite{Ranft95b}. The HADRIN model used
by DPMJET for hadron--hadron collisions below 5 GeV was compared to data
down to several hundred MeV\cite{Haenssgen86,Haenssgen84a}.
  DPMJET--II.5  works 
 for nucleus--nucleus interactions above 5~GeV
per nucleon,
 but was never compared
 to experimental data below 60 GeV.

DPMJET--II.5 is able to run up to energies of
approximately $10^{21}$ eV in the lab system (or $\sqrt s$ =
2000 TeV).

%
 \newpage


\section{Comparing DPMJET--II.5 to data}


\subsection{Comparing to data in hadron--hadron collisions}

Before a model like DPMJET--II.5 can be used for anything, we
have to demonstrate, that the model describes well enough the
available data up to the highest energies. The parameters of the
model as given in Table 1 are different from the
parameters used in the previous version of the model, therefore,
this agreement to the data is not trivial, even if the previous
DPMJET version was quite successful in this respect.

{\bf Table 4.}
Comparison of average multiplicities of produced hadrons in
proton-- proton collisions at 200 GeV. The experimental data are
from Ref.\cite{marek}.
\vskip 5mm
\begintable
Particle   | DPMJET--II.5 | Exp.\cite{marek} ~\cr
 $n_{ch}$  | 7.69  | 7.69 $\pm$ 0.06\cr
 $n_-$     | 2.85  | 2.85 $\pm$ 0.03\cr
 p         | 1.24  | 1.34 $\pm$ 0.15\cr
 n         | 0.68  | 0.61 $\pm$ 0.30\cr
 $\pi^+$   | 3.25  | 3.22 $\pm$ 0.12\cr
 $\pi^-$   | 2.58  | 2.62 $\pm$ 0.06\cr
 $\pi^0$   | 3.55  | 3.34 $\pm$ 0.24\cr
 $K^+$     | 0.30  | 0.28 $\pm$ 0.06\cr
 $K^-$     | 0.19  | 0.18 $\pm$ 0.05\cr
 $K^0_S$   | 0.23  | 0.17 $\pm$ 0.01\cr
 $\Lambda$  |0.14   |0.096 $\pm$  0.01\cr
 $\bar\Lambda$  |0.0178 |0.0136 $\pm$ 0.004 \cr
 $\rho^0$  |0.48   |0.33 $\pm$  0.06\cr
 $\bar p$  | 0.048 | 0.05 $\pm$ 0.02 \endtable
\vskip 10mm 

\clearpage

We start with looking at data, which prove, that the average
multiplicity and the particle content are well described.
In
Table 4 we compare to the very well known multiplicities of
secondary hadrons in 200 GeV p--p collisions 
from Ref.\cite{marek}.
In
Table 5 we compare to  multiplicities of
secondary hadrons  and hadron resonances 
in $\sqrt s$ = 27.5 GeV p--p collisions 
from Ref.\cite{Aguilar91a}.


{\bf Table 5.}
Comparison of average multiplicities of produced hadrons  and
hadron resonances in
proton-- proton collisions at $\sqrt s$ = 27.5  GeV.
The experimental data are
from Ref.\cite{Aguilar91a}.
\vskip 5mm
\begintable
Particle   | DPMJET--II.5 | Exp.\cite{Aguilar91a} ~\cr
 p         | 1.28  | 1.20 $\pm$ 0.097 $\pm$ 0.022\cr
 $\pi^+$   | 3.52  | 4.10 $\pm$ 0.11 $\pm$ 0.15\cr
 $\pi^-$   | 2.87  | 3.34 $\pm$ 0.08 $\pm$ 0.12\cr
 $\pi^0$   | 3.88  | 3.87 $\pm$ 0.12 $\pm$ 0.16\cr
 $K^+$     | 0.34  | 0.331 $\pm$ 0.016 $\pm$ 0.007\cr
 $K^-$     | 0.24  | 0.224 $\pm$ 0.011 $\pm$ 0.004\cr
 $K^0_S$   | 0.27  | 0.232 $\pm$ 0.011 $\pm$ 0.004\cr
 $\Lambda$  |0.156  |0.125 $\pm$  0.008 $\pm$ 0.008\cr
 $\bar\Lambda$  |0.026  |0.020 $\pm$ 0.004  $\pm$ 0.0008\cr
 $\Sigma^+$  |0.050  |0.048 $\pm$  0.015 $\pm$ 0.004\cr
 $\Sigma^-$  |0.016  |0.0128 $\pm$  0.0061 $\pm$ 0.0032\cr
 $\Delta^{++}$  |0.24  |0.218 $\pm$  0.0031 $\pm$ 0.013\cr
 $\Delta^{0}$  |0.23  |0.141 $\pm$  0.0098 $\pm$ 0.0089\cr
 $\bar\Delta^{++}$  |0.005 |0.013 $\pm$  0.0049 $\pm$ 0.0049\cr
 $\bar\Delta^{0}$  |0.006 |0.0336 $\pm$  0.008 $\pm$ 0.0006\cr
 $\rho^0$  |0.54   |0.385 $\pm$  0.018 $\pm$ 0.038\cr
 $\rho^+$  |0.54   |0.552 $\pm$  0.083 $\pm$ 0.046\cr
 $\rho^-$  |0.43   |0.355 $\pm$  0.058 $\pm$ 0.033\cr
 $\omega$  |0.50   |0.390 $\pm$  0.024 $\pm$ 0.002\cr
 $\eta$  |0.55   |0.30 $\pm$  0.02 $\pm$ 0.054\cr
 $K^{*+}$     | 0.18  | 0.132 $\pm$ 0.016 $\pm$ 0.002\cr
 $K^{*-}$     | 0.11 | 0.088 $\pm$ 0.012 $\pm$ 0.001\cr
 $K^{*0}$   | 0.15  | 0.119 $\pm$ 0.021 $\pm$ 0.002\cr
 $\bar K^{*0}$   | 0.11 | 0.0903 $\pm$ 0.016 $\pm$ 0.001\cr
 $\phi$  |0.029  |0.019 $\pm$  0.0018 $\pm$ 0.054\cr
 $\bar p$  | 0.066 | 0.063 $\pm$ 0.002 $\pm$ 0.001 \endtable
\vskip 10mm

\begin{figure}[thb]
\begin{center}
 \psfig{figure=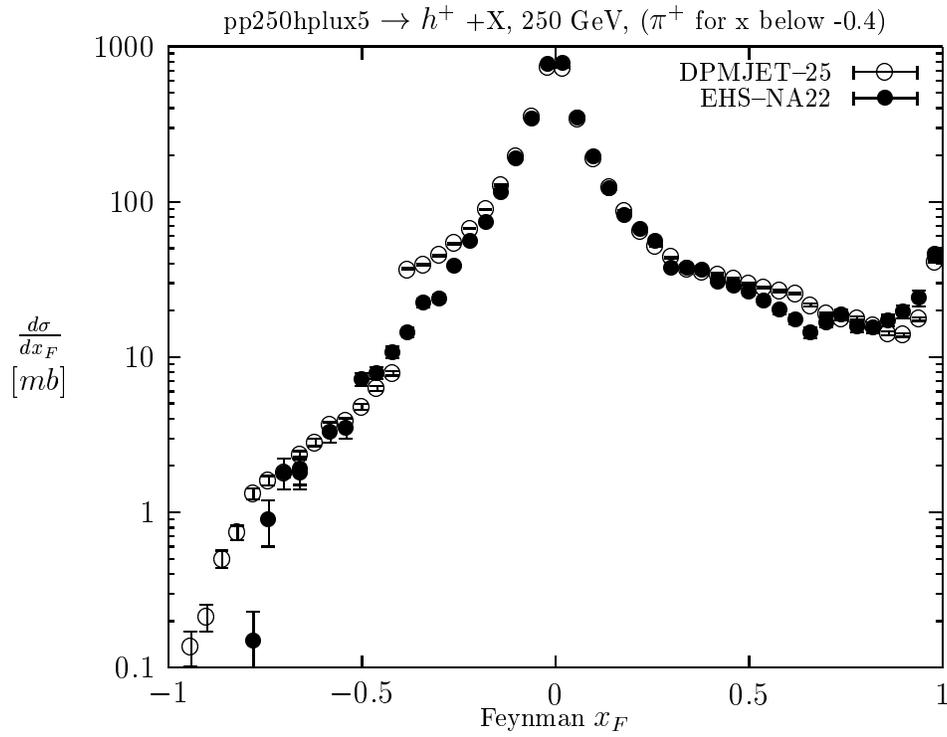}
\end{center}
\vspace*{-3mm}
\caption{Comparison of Feynman--x distributions of positively charged
hadrons
 produced in proton--proton collisions at 250 GeV. The
experimental data are from the EHS--NA22 Collaboration
\protect\cite{Adamus88a}.
\protect\label{pp250hplux}
}
\end{figure}
\begin{figure}[thb]
\begin{center}
 \psfig{figure=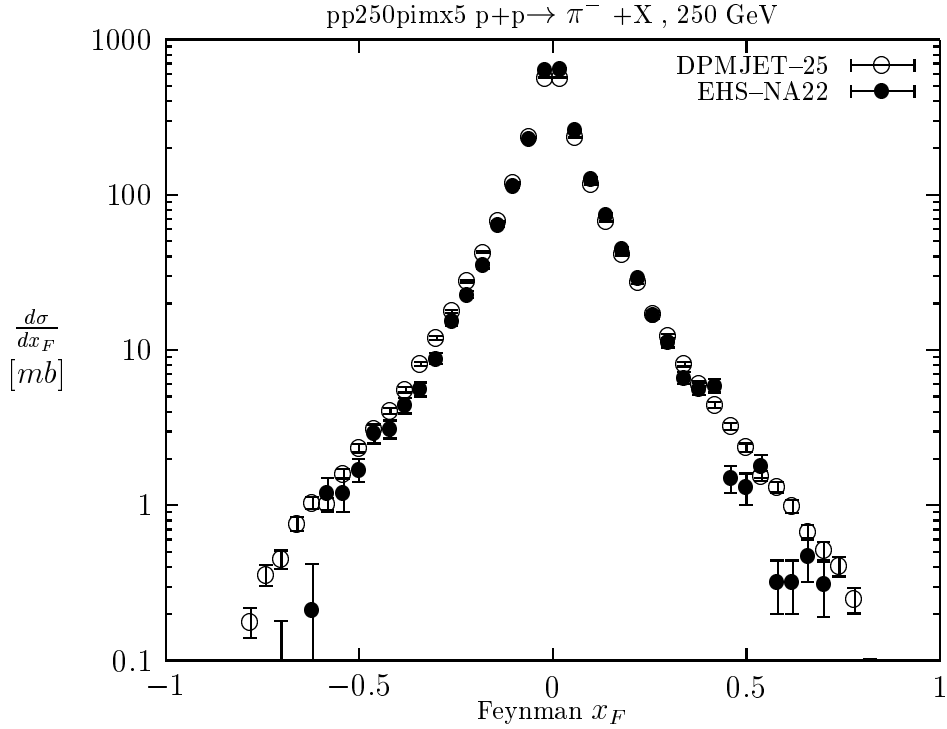}
\end{center}
\vspace*{-3mm}
\caption{Comparison of Feynman--x distributions of $\pi^-$
mesons produced in proton--proton collisions at 250 GeV. The
experimental data are from the EHS--NA22 Collaboration
\protect\cite{Adamus88a}.
\protect\label{pp250pimx}
}
\end{figure}
%
\begin{figure}[thb] \centering
 \psfig{figure=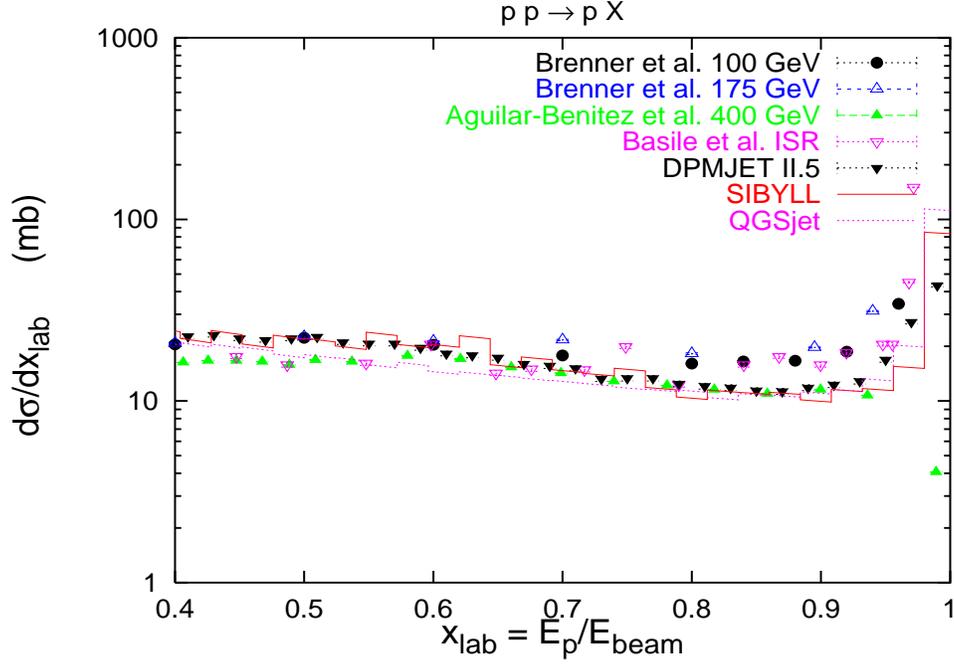,width=12.0cm,height=9.0cm}
\vspace*{8mm}
\caption{
 $x_{lab}$ cross sections in p--p collisions. We compare data at
 different energies \protect\cite{Brenner82,Aguilar91a,Basile84} shown as
 symbols with
 DPMJET--II.5 at \protect$p_{lab}$ = 200 GeV/c shown as solid triangles.
 The data are also compared to
 the SIBYLL \protect\cite{Fletcher94} and QGSJET 
 \protect\cite{QGSJET} models shown as
 histograms.
\protect\label{lead-bar-low1}
}
\end{figure}
%
%
\begin{figure}[thb] \centering
 \psfig{figure=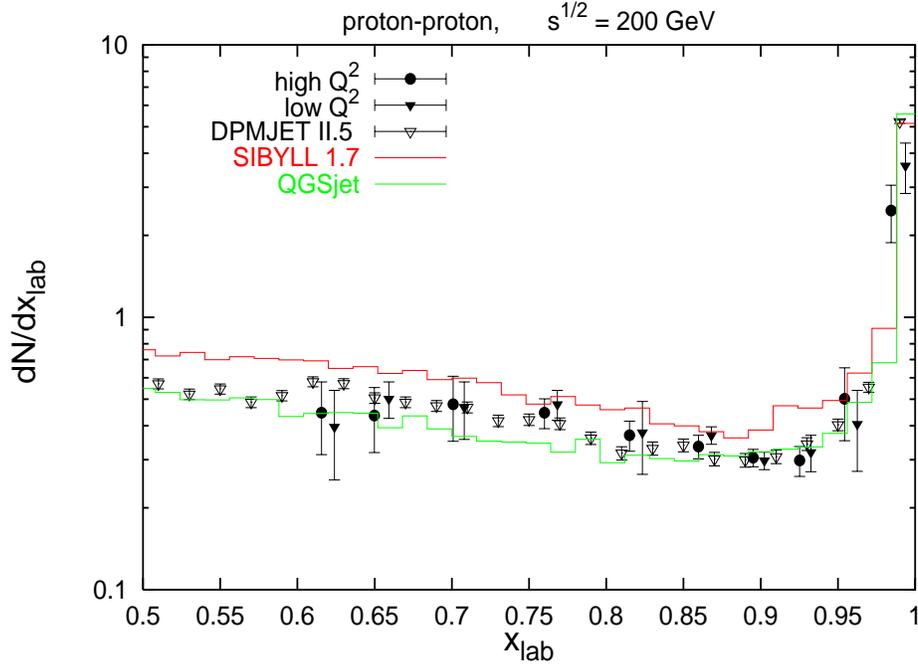,width=12.0cm,height=9.0cm}
\vspace*{8mm}
\caption{
Energy fraction \protect$x_{lab}$ carried by the leading proton. The data are
photoproduction and DIS measurements at $\sqrt s$ = 200 GeV
\protect\cite{Garfagnini98,Schmidke98} 
shown as symbols compared to DPMJET--II.5 for p--p
collisions at  $\sqrt s$ = 200 GeV shown as open triangles. 
 The data are also compared to
 the SIBYLL \protect\cite{Fletcher94} and QGSJET 
 \protect\cite{QGSJET} models shown as
 histograms.
\protect\label{lead-pro2}
}
\end{figure}
%

In Fig.  \ref{pp250hplux}  and \ref{pp250pimx} 
we compare the model with NA-22 data on
the Feynman--$x$ distribution of positively charged hadrons and 
of $\pi^-$ produced in 250 GeV
$pp$ collisions. In Fig. \ref{pp250hplux} we observe in particular
(compared to previous DPMJET versions)  a
better agreement of the model predictions with the forward production of
protons.  

Distributions of leading protons are compared more directly 
to data in the following two Figs. \ref{lead-bar-low1} 
and \ref{lead-pro2}. The  Figs. are adopted 
from  talks of Engel \cite{Engel99a,Engel99b}. But of course, we present
now the DPMJET--II.5 results.
 The leading 
particle
production is very important for the Cosmic Ray cascade simulation.

 In  Fig. \ref{lead-bar-low1} we compare
 $x_{lab}$ cross sections in p--p collisions measured at
 different energies \cite{Brenner82,Aguilar91a,Basile84} with
 DPMJET--II.5 results at $p_{lab}$ = 200 GeV/c. 
 The DPMJET cross sections change
 only very slowly with energy. The obtained agreement with DPMJET--II.5 
 to the data is much better than with former DPMJET versions see
 \cite{Engel99a}.
 The data are also compared to
 the SIBYLL \cite{Fletcher94} and QGSJET \cite{QGSJET} models.

 In Fig.  \ref{lead-pro2} we compare 
the  energy fraction $x_{lab}$ carried by the leading proton. 
The data are
photoproduction and DIS measurements from the HERA Collider 
at $\sqrt s$ = 200 GeV
\cite{Garfagnini98,Schmidke98}. We compare to DPMJET--II.5 for p--p
collisions at  $\sqrt s$ = 200 GeV. The forward production of leading
protons is not expected to depend strongly on the reaction channel. It
is found again, that DPMJET--II.5 agrees much better to the data than
older DPMJET versions, see \cite{Engel99b}. This
comparison demonstrates, that the leading proton distribution at $\sqrt
s$ = 200 GeV is still rather flat like at the much lower energies in
Fig. \ref{lead-bar-low1}.
At $\sqrt s$ = 200 GeV  
both versions of DPMJET--II.5 
with GSQBS and USQBS and with only GSQBS, (discussed in
Section II) lead still to very similar $x_{lab}$ distributions of
secondary protons. Unfortunately therefore, these data cannot be used to
discriminate between the two DPMJET--II.5 versions.
 The data are also compared to
 the SIBYLL \cite{Fletcher94} and QGSJET \cite{QGSJET} models.
 


\begin{figure}[thb]
\begin{center}
 \psfig{figure=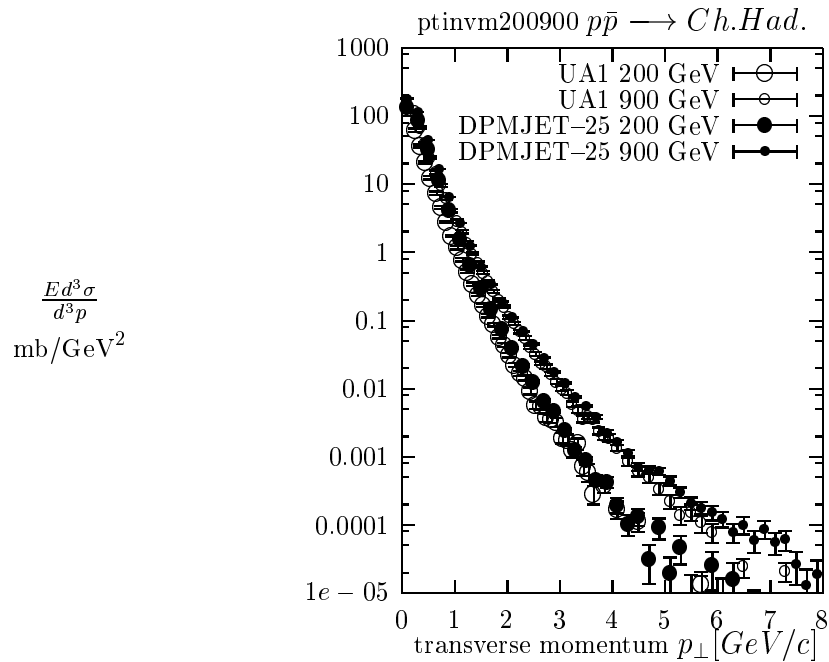}
\end{center}
\vspace*{-3mm}
\caption{Comparison with transverse momentum cross sections at
$\sqrt s$ = 0.2 and -.9 TeV with
collider data from the UA--1 Experiment \protect\cite{Albajar90}.
\label{ptinvm200900}
}
\end{figure}
\begin{figure}[thb]
\begin{center}
 \psfig{figure=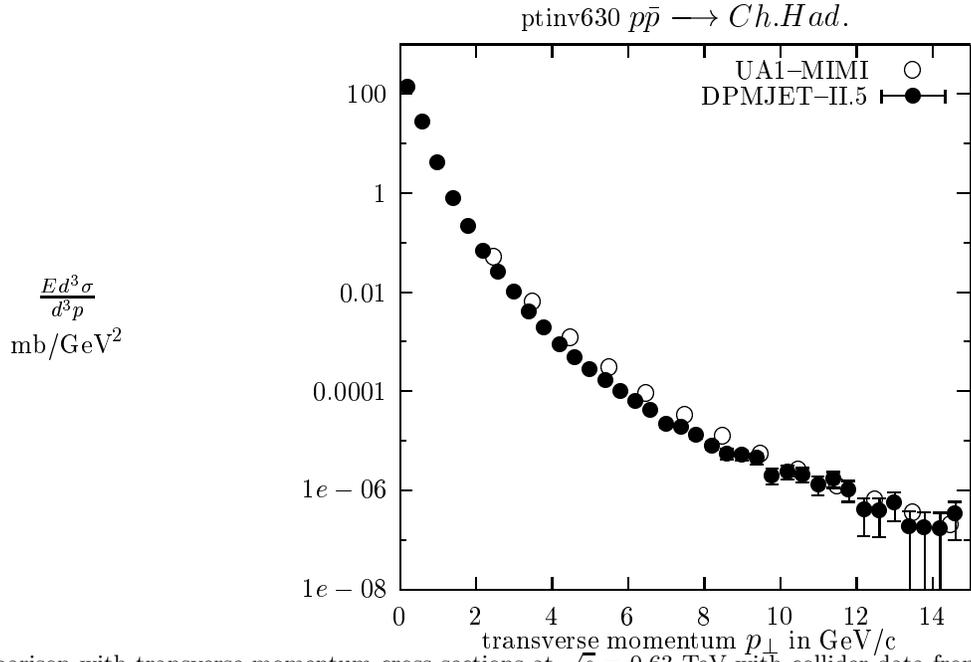}
\end{center}
\vspace*{-3mm}
\caption{Comparison with transverse momentum cross sections at
$\sqrt s$ = 0.63  TeV with
collider data from the UA--1 MIMI Collaboration 
\protect\cite{Bocquet96a}. The experimental data are represented
by the parametrization given by the Experiment.
\label{ptinv630}
}
\end{figure}
\begin{figure}[thb]
\begin{center}
 \psfig{figure=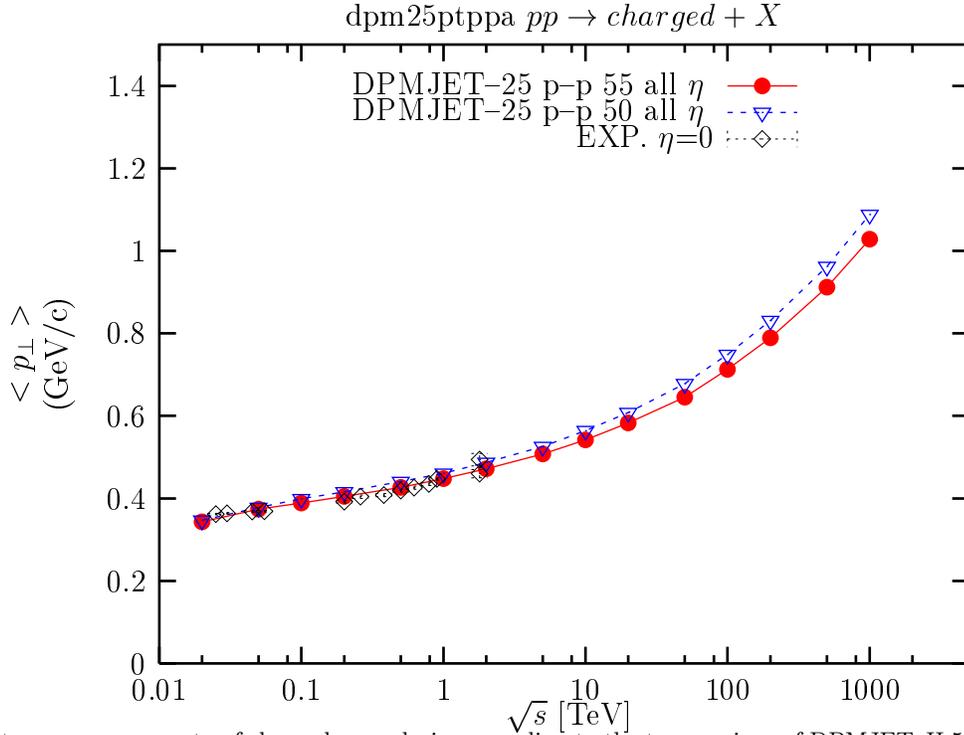}
\end{center}
\vspace*{-3mm}
\caption{Average transverse momenta of charged secondaries according to
the two versions of DPMJET--II.5 
produced in $pp$ collisions  in the full
 pseudorapidity interval as function of the center of
mass energy $\sqrt s$. We compare to data collected
by the UA--1 Collaboration \protect\cite{Albajar90} which refer to the
central pseudorapidity range and are therefore expected to be slightly
larger than in the full range.
\label{dpm222ptav}
}
\end{figure}


 Hadron transverse momentum distributions were measured by the
 UA--1 Collaboration
\cite{Albajar90}. In Fig.\ref{ptinvm200900} we compare the
distributions at $\sqrt s$ = 200 GeV and 900 GeV with
DPMJET--II.5. The transverse momentum distribution up to larger
$p_{\perp}$ was  determined by the UA--1--MIMI Collaboration
\cite{Bocquet96a}. In Fig.\ref{ptinv630} we compare  the model
results with the parametrization of the data given by this
experiment.

In Fig.\ref{dpm222ptav} we compare average transverse momenta as
obtained from the two versions discussed in Section II of 
DPMJET--II.5 as function of the cms energy $\sqrt
s$ with data collected by the UA--1 Collaboration. This plot
gives at the same time the DPMJET--II.5 predictions for the average
transverse momenta up to $\sqrt s$ = 1000 TeV. We find, that the average
transverse momenta in both versions of the model agree rather well to
each other. Average
transverse momenta depend on the longitudinal variables, they
differ if one chooses different pseudorapidity ranges, the data given
for the central pseudorapidity range are expected to be slightly larger
than the calculation given for the full range. 



\begin{figure}[thb]
\begin{center}
 \psfig{figure=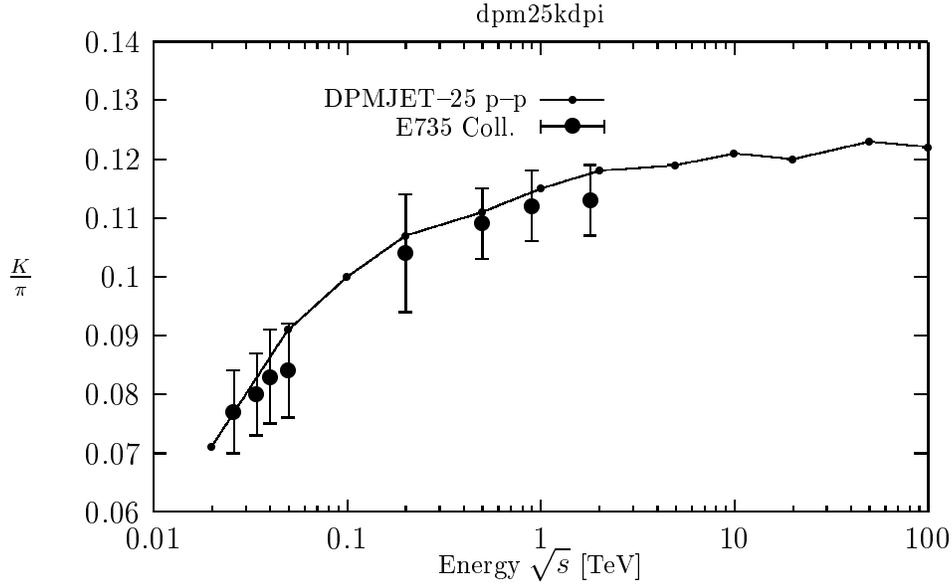}
\end{center}
\vspace*{-3mm}
\caption{K/$\pi$ ratios in $pp$ or  $p\bar p$ collisions as function of
the cms energy $\sqrt s$. The DPMJET--II.5 calculation is
compared with data collected from the E735 Collaboration at
Fermilab \protect\cite{Alexopoulos92}.
\label{dpm222kdpi}
}
\end{figure}

With rising energy the fraction of strange hadrons rises slowly.
In Fig.\ref{dpm222kdpi} the K/$\pi$ ratio according to DPMJET--II.5
for $ pp$ or $p\bar p$ collisions is compared to data from the
E735 Collaboration
\cite{Alexopoulos92}.
%
 \clearpage

 
\subsection{Comparing to data in hadron--nucleus and
nucleus--nucleus collisions}

We now turn to collisions with nuclei.

In Table 6 we compare average 
multiplicities of negatively charged hadrons calculated with
DPMJET--II.5 in
hadron--hadron, minimum bias hadron--nucleus and central
nucleus--nucleus collisions with experimental data.

{\bf Table 6.}
Comparison of average multiplicities of produced negatively charged 
hadrons in
proton--proton, minimum bias proton--nucleus and central
nucleus--nucleus collisions at 200 GeV.
\vskip 5mm
\begintable
Collision   | DPMJET--II.5 | Exp.|Reference ~\cr
 p--p     | 2.85  | 2.85 $\pm$ 0.03 | \cite{marek}  \cr
 p--S         | 5.10  |5.0  $\pm$0.2 | \cite{NA35}  \cr
 p--Ar         |5.30   |5.39  $\pm$0.17 | \cite{DeMarzo82}  \cr
 p--Ag         | 6.18  |6.2  $\pm$0.2 | \cite{Brick89}  \cr
 p--Xe         | 6.43  |6.84  $\pm$0.13 | \cite{DeMarzo82}  \cr
 p--Au         | 6.81  |7.0  $\pm$0.4 | \cite{Brick89}  \cr
 p--Au         | 6.81  |7.3  $\pm$0.3 | \cite{Baechler91}  \cr
 S--S central         |103   |98  $\pm$3 | \cite{Alber98}  \cr
 S--Ag central         |174   |186  $\pm$11 | \cite{Alber98}  \cr
 S--Au central         | 202  |225  $\pm$12 | \cite{Alber98}  \endtable
\vskip 10mm 
{\bf Table 7.}
Comparison of average multiplicities of produced strange 
hadrons in
 central S--S and S--Ag
 collisions at 200 GeV.
 For DPMJET--II.5 we give  the results for the models without and with
 secondary interactions  of comovers 
 and it is assumed that $\Sigma^0$ and
 $\bar \Sigma^0$ have decayed. The experimental data are from the NA35
 Collaboration \cite{Seyboth97}.
\vskip 5mm
\begintable
Reaction|Particle | DPMJET--II.5 |DPMJET--II.5 with sec.int.| Exp.\cite{Seyboth97} ~\cr
S--S | $\Lambda$  | 5.7  |7.3 |9.4 $\pm$ 1.0   \cr
S--S | $\bar \Lambda$  |1.13   |1.22 |2.6 $\pm$ 0.3   \cr
S--S |$K^0_S$  |8.7   |9.1 |10.5 $\pm$ 1.7   \cr
S--S |$K^+$  | 10.6  |11.4 |12.5 $\pm$ 0.4   \cr
S--S |$K^-$  | 6.8  |6.9 |6.9 $\pm$ 0.4   \cr
S--Ag | $\Lambda$  |9.6   |13.1 |15.2 $\pm$1.2    \cr
S--Ag | $\bar \Lambda$  | 1.8  |2.0 |2.6 $\pm$ 0.3   \cr
S--Ag | $K^0_S$  | 14.6  |15.5 |15.5 $\pm$1.5    \cr
S--Ag | $K^+$  | 17.8  |19.7 |17.4 $\pm$ 1.0   \cr
S--Ag | $K^-$  | 11.4  |11.4 |9.6 $\pm$ 1.0   \endtable
\vskip 10mm 


In Tables 7 and 8 we give the  average multiplicities 
of produced strange hadrons in central S--S, S--Ag and Pb-Pb
 collisions at 200 GeV (158 GeV for Pb--Pb).
 We give the  DPMJET--II.5  
 results for the models without and with
 secondary interactions and it is assumed, that $\Sigma^0$ and
 $\bar \Sigma^0$ have decayed. The experimental data for central S--S
and  S--Ag collisions are from the NA35
 Collaboration \cite{Seyboth97}. There is some doubt, that the simple
 model for secondary interactions in DPMJET--II.5 gives correct results
 in a situation with such high produced hadron density like in the
 central Pb--Pb collisions. 
 
 In Table 8 we give the results for central 
 Pb--Pb collisions also for the RHIC and LHC energies and we include
 charged multiplicities and and central plateau heights in the Table.
 At the LHC DPMJET--II.5 gives a slightly lower plateau then
 DPMJET--II.2 (for which the plateau was given in \cite{ALICETP}). The
 central plateaus at RHIC and LHC energies 
 given in Table 8 agree however
 very well with the ones given by Capella in \cite{Capella99b}. 

{\bf Table 8.}
Comparison of average multiplicities of produced strange 
hadrons in
 central Pb--Pb  
 collisions at 158 GeV and at the energies of RHIC and the LHC.
 For DPMJET--II.5 we give only at 158 GeV 
 the results for the models without and with
 secondary interactions of comovers 
 and it is assumed that $\Sigma^0$ and
 $\bar \Sigma^0$ have decayed. 
\vskip 5mm
\begintable
Particle | 158 GeV DPMJET--II.5 |158 GeV DPMJET--II.5 with sec.int.|RHIC |LHC ~\cr
 $\Lambda$  |51   |84 |107 |242  \cr
 $\bar \Lambda$  |11.9  |14.6 |54 | 181   \cr
 $K^0_S$  | 70  |79 |359 |1306    \cr
 $K^+$  | 86  |104|390 |1359   \cr
$K^-$  | 54  |56 |341 | 1302   \cr
$\Xi^-$  | 3.81  |3.80 |10.0 | 24.2   \cr
$\bar \Xi^+$  |1.67   |1.64 |6.0 | 19.1   \cr
$\Omega^-$  | 0.24  |0.24 |0.9 | 2.2   \cr
$\bar \Omega^+$  |0.18   |0.17 |0.7 | 1.9   \cr
$(dN^{ch}/dy^*)_{y^*=0}$  |525   |516 |1282 |2801   \cr
 Fraction of cent. coll. (\%) | 5  |5 |3 | 4  \cr
$h^-$  | 828  |807 |3560 |12581   \endtable
\vskip 10mm 

\begin{figure}[thb]
\begin{center}
 \psfig{figure=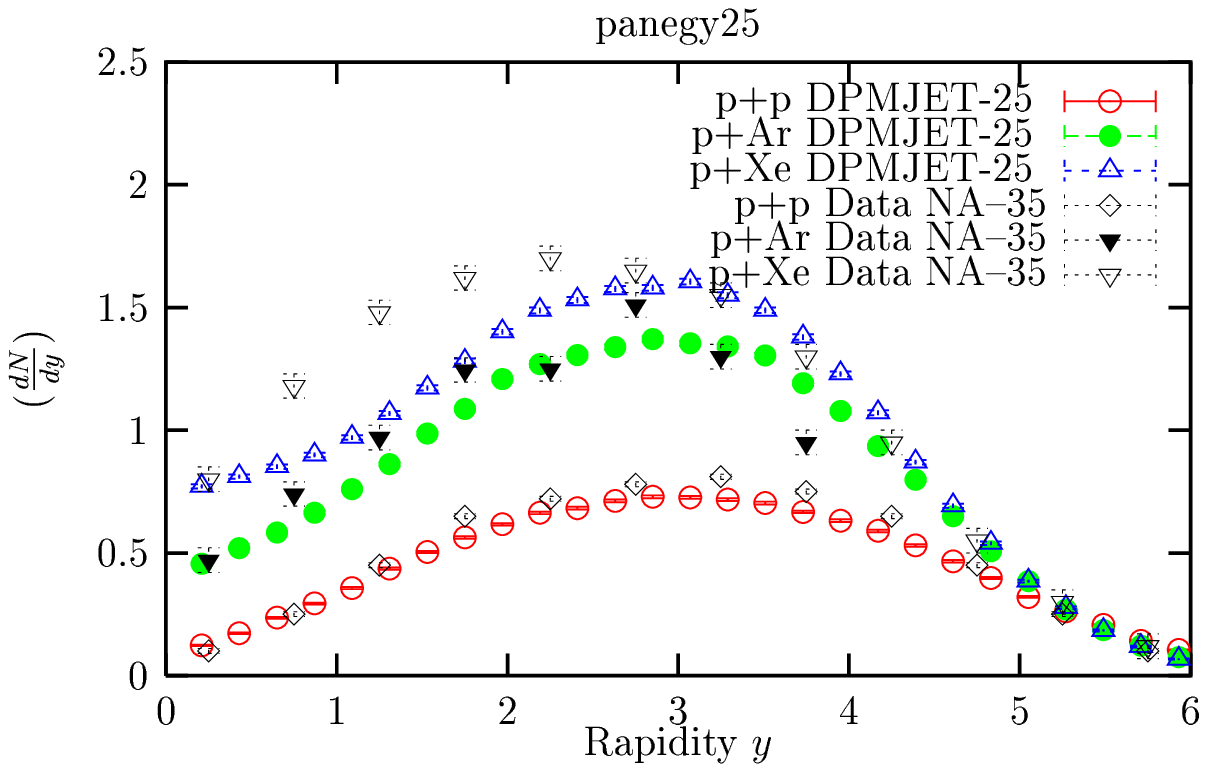}
\end{center}
\begin{center}
 \psfig{figure=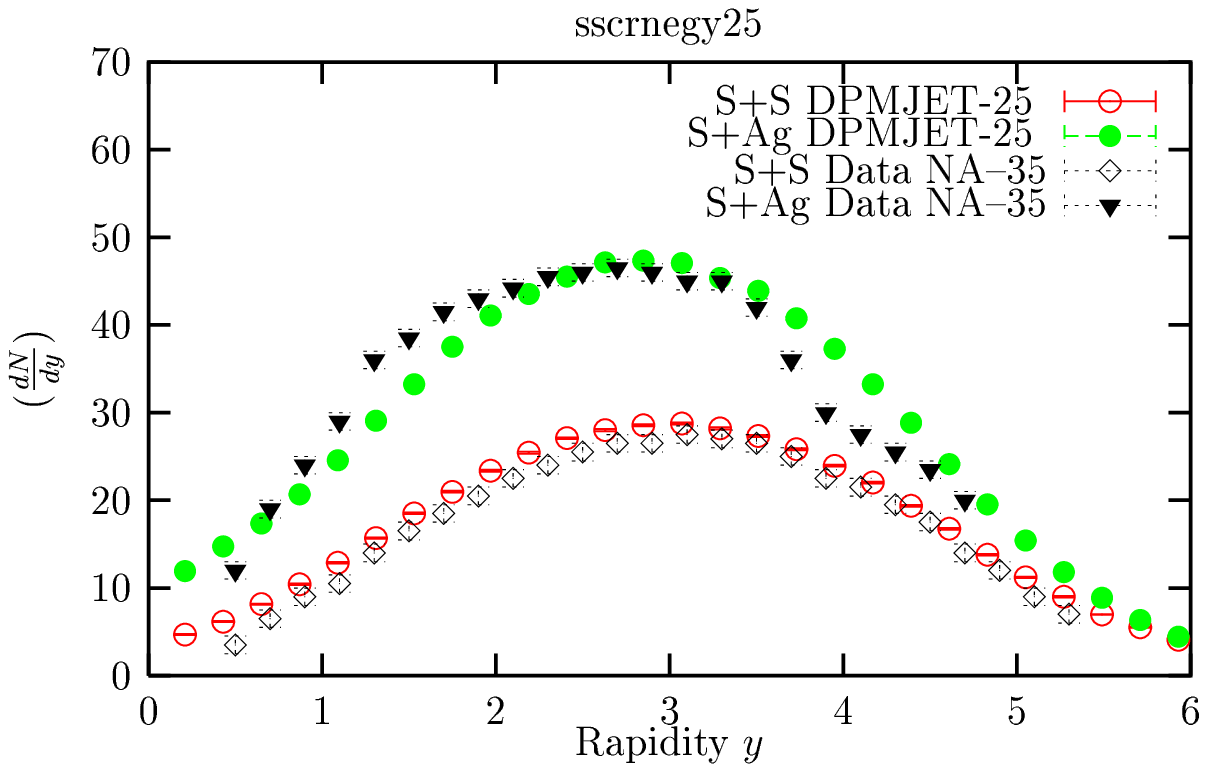}
\end{center}
\vspace*{-3mm}
\caption{{\bf (a) }Negatively charged particle rapidity distribution for p--p, 
p--Ar and p--Xe
interactions. The DPMJET--II.5 results are compared with data
\protect\cite{Alber98}.
{\bf (b) }Rapidity distribution of negatively charged hadrons in
central S--S ans S--Ag collisions. The results of DPMJET--II.5
are compared with data from the NA--35 Collaboration
\protect\cite{Alber98}.
\label{rappar}
}
\end{figure}

In Fig. \ref{rappar}.a the
comparison is with the rapidity distribution of negatively 
charged hadrons
in p--p, p-Ar and p--Xe 
collisions at 200 GeV and in Fig. \ref{rappar}.b we
compare to rapidity distributions of negatively charged hadrons
in central S--S and S--Ag collisions.
 \clearpage

 \clearpage

\begin{figure}[thb]
\begin{center}
 \psfig{figure=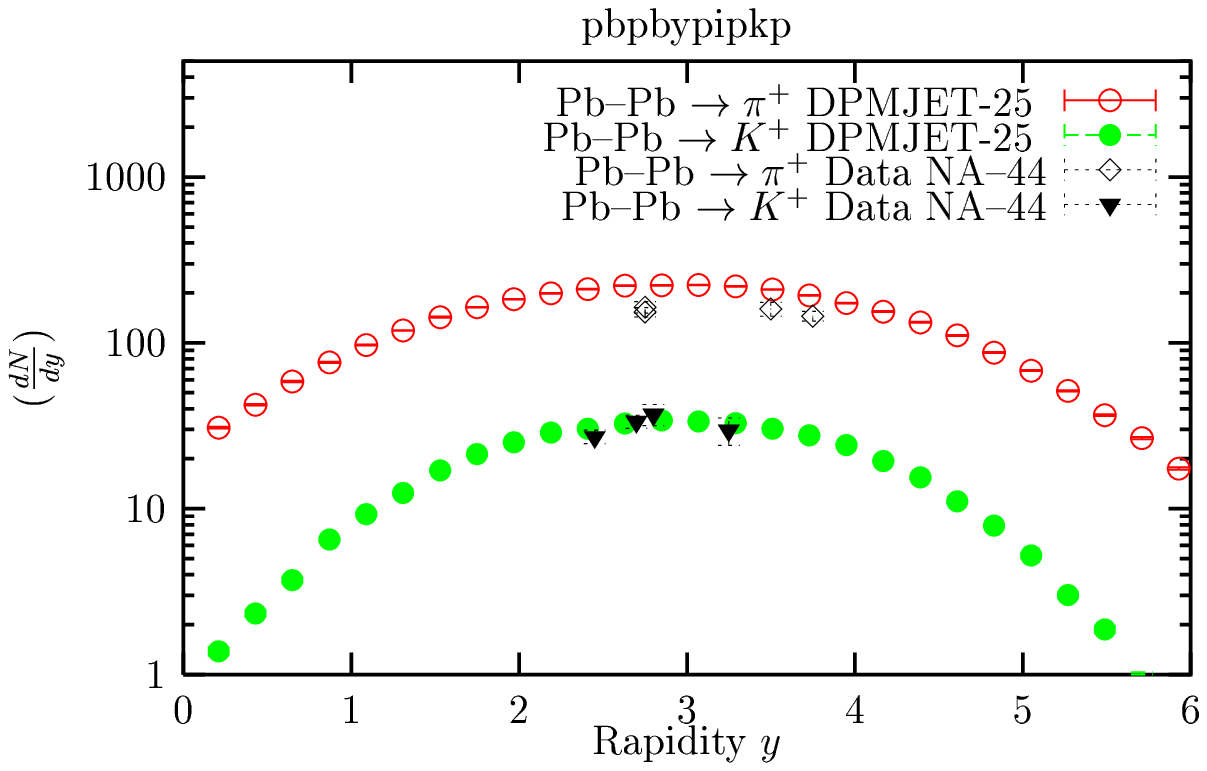}
\end{center}
\begin{center}
 \psfig{figure=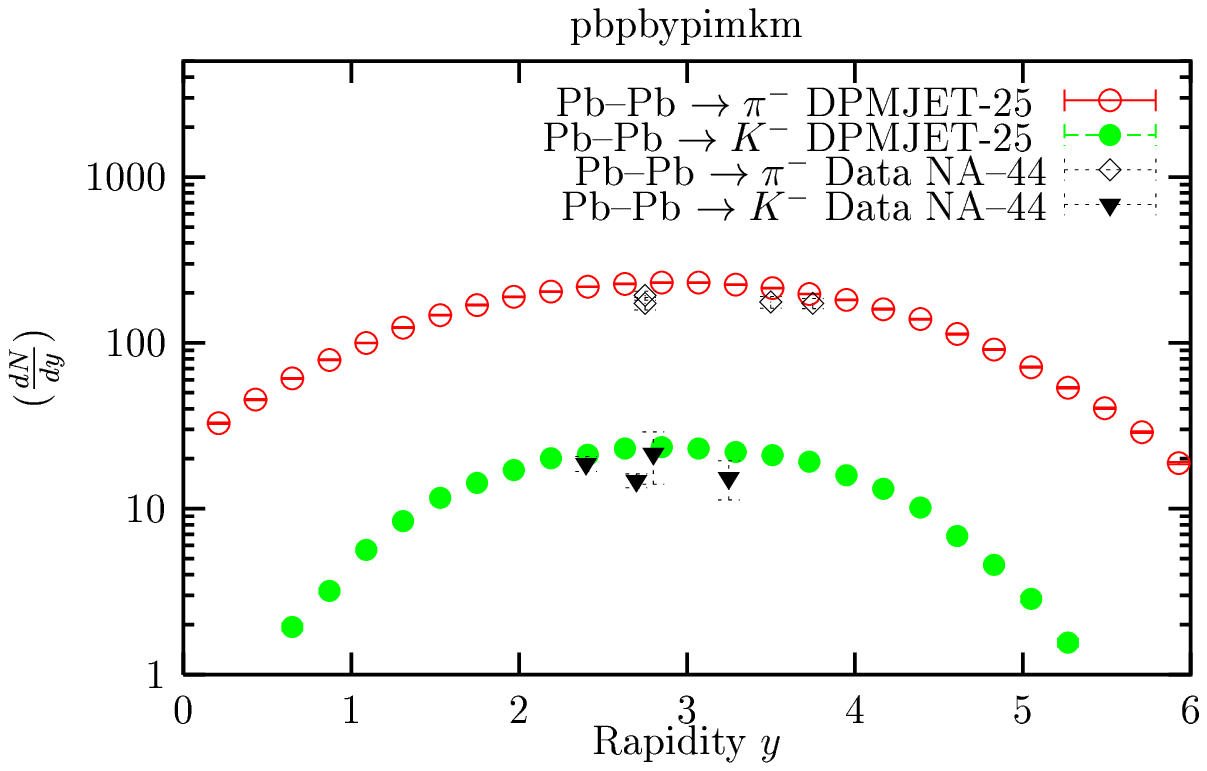}
\end{center}
\vspace*{-3mm}
\caption{{\bf (a) }Rapidity distributions of $\pi^+$ and $K^+$ mesons
produced in central Pb--Pb collisions. The DPMJET--II.5 distributions
are compared with experimental data from the NA--44 Collaboration
\protect\cite{Bearden99}.
{\bf (b) }
Rapidity distributions of $\pi^-$ and $K^-$ mesons
produced in central Pb--Pb collisions. The DPMJET--II.5 distributions
are compared with experimental data from the NA--44 Collaboration
\protect\cite{Bearden99}.
\label{pbpbypipkp}
}
\end{figure}

In Fig. \ref{pbpbypipkp} we compare rapidity distributions of $\pi^+$,
$\pi^-$, $K^+$ and $K^-$ mesons produced in central (4\%) Pb--Pb
collisions at 158 AGeV/c. The experimental data, mainly in the central
rapidity region are from the NA--44 Collaboration \cite{Bearden99}. The
agreement of model and data is encouraging, however not perfect.

 \clearpage

 
\subsection{Baryon stopping}

\begin{figure}[thb]
\begin{center}
 \psfig{figure=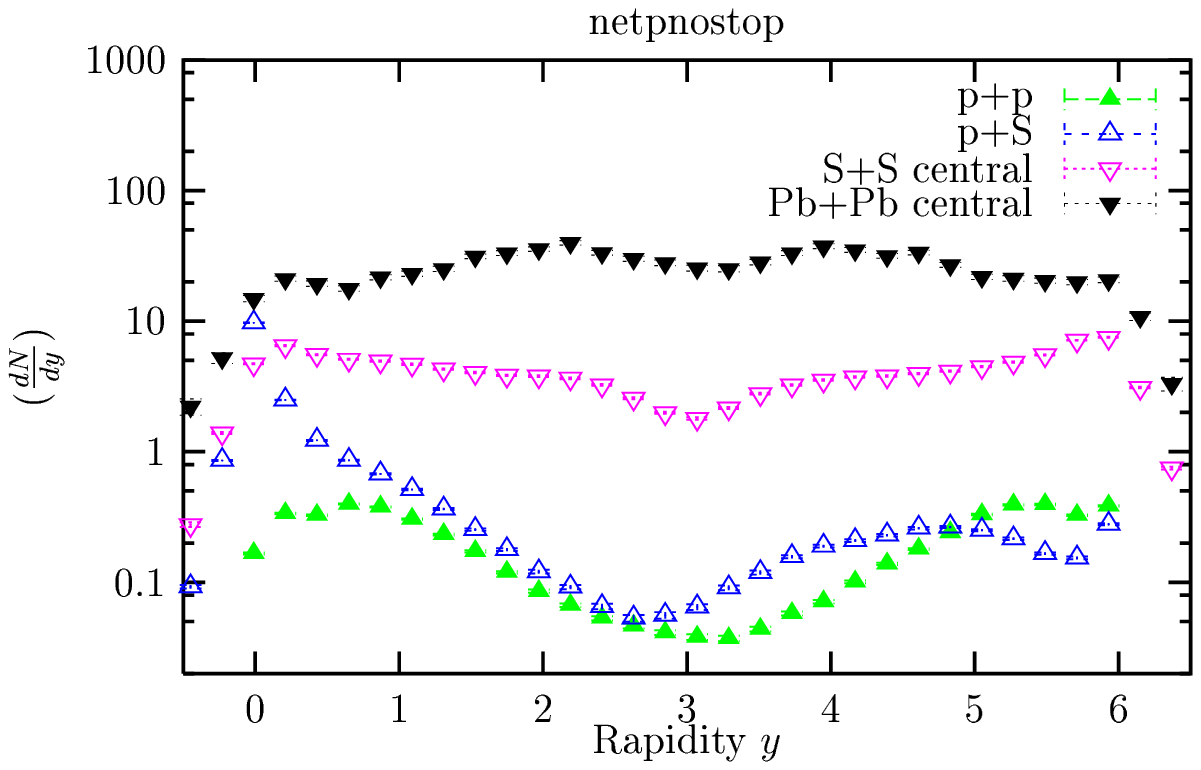}
\end{center}
\begin{center}
 \psfig{figure=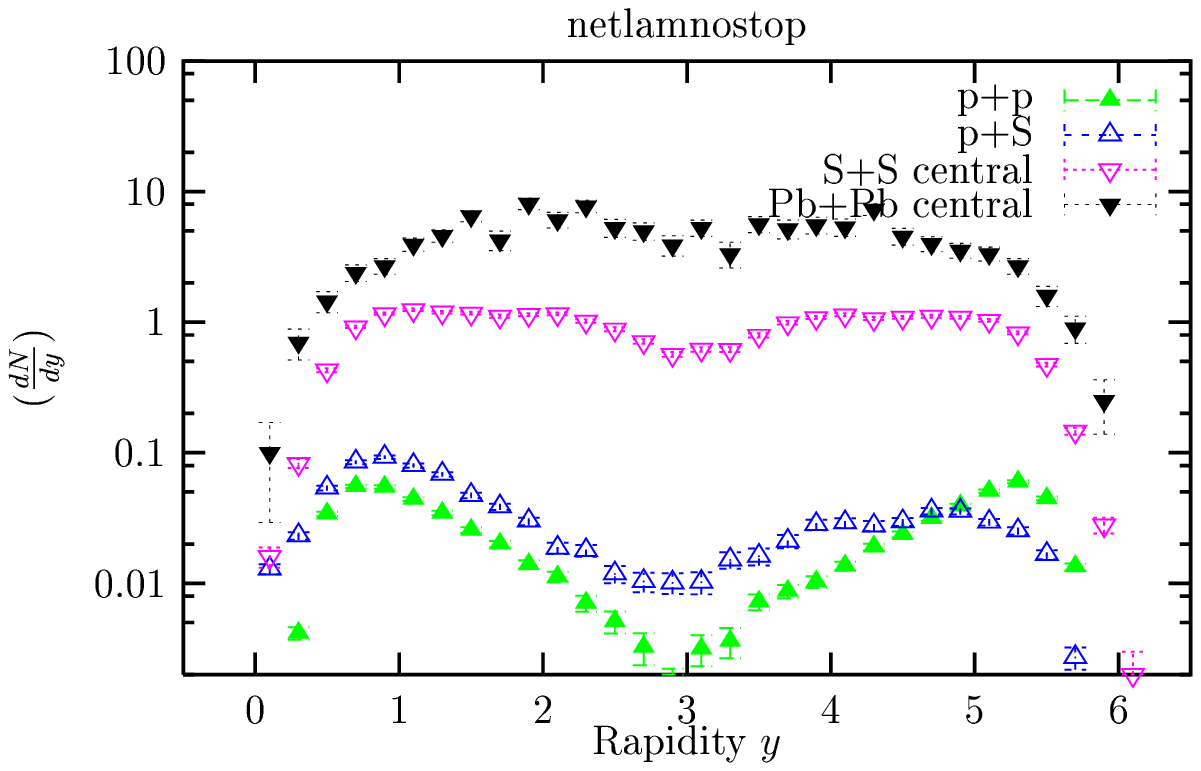}
\end{center}
\vspace*{-3mm}
\caption{{\bf (a) }Net proton ($p-\bar p$) rapidity distribution in p--p,
p--S, central S--S and central Pb--Pb. 
collisions. Calculated with DPMJET--II.5 without the new diagrams
modifying baryon stopping.
{\bf (b) }Net $\Lambda$ ($\Lambda-\bar \Lambda$) rapidity distribution in p--p,
p--S, central S--S and central Pb--Pb. 
collisions. Calculated with DPMJET--II.5 without the new  GSQBS and
USQBS diagrams
modifying baryon stopping.
\label{netpnostop}
}
\end{figure}

We present first the DPMJET predictions at $p_{lab}$ = 200 GeV/c 
 for net baryon rapidity
distributions in the original model without the new GSQBS and USQBS 
diagrams. In Fig.
\ref{netpnostop}.a we present the leading 
net proton ($p-\bar p$) rapidity
distribution $dN_p/dy -dN_{\bar p}/dy$ in p--p,
 p--S, central S--S and central Pb--Pb collisions.
 In Fig.
\ref{netpnostop}.b we present the 
net  $\Lambda$ ($\Lambda-\bar \Lambda$)  rapidity
distribution $dN_{\Lambda}/dy - dN_{\bar \Lambda}/dy$ in p--p,
 p--S, central S--S and central Pb--Pb collisions. In p--p collisions,
 which at the given collision energy are hardly modified by the new
GSQBS and USQBS  diagrams, 
we observe a dip at central rapidity. This dip is also
 present in DPMJET without the new GSQBS and USQBS  diagrams 
 in p--S and central A--A collisions. This disagrees to the data
 presented in the next Figures.

 \clearpage

\begin{figure}[thb]
\begin{center}
 \psfig{figure=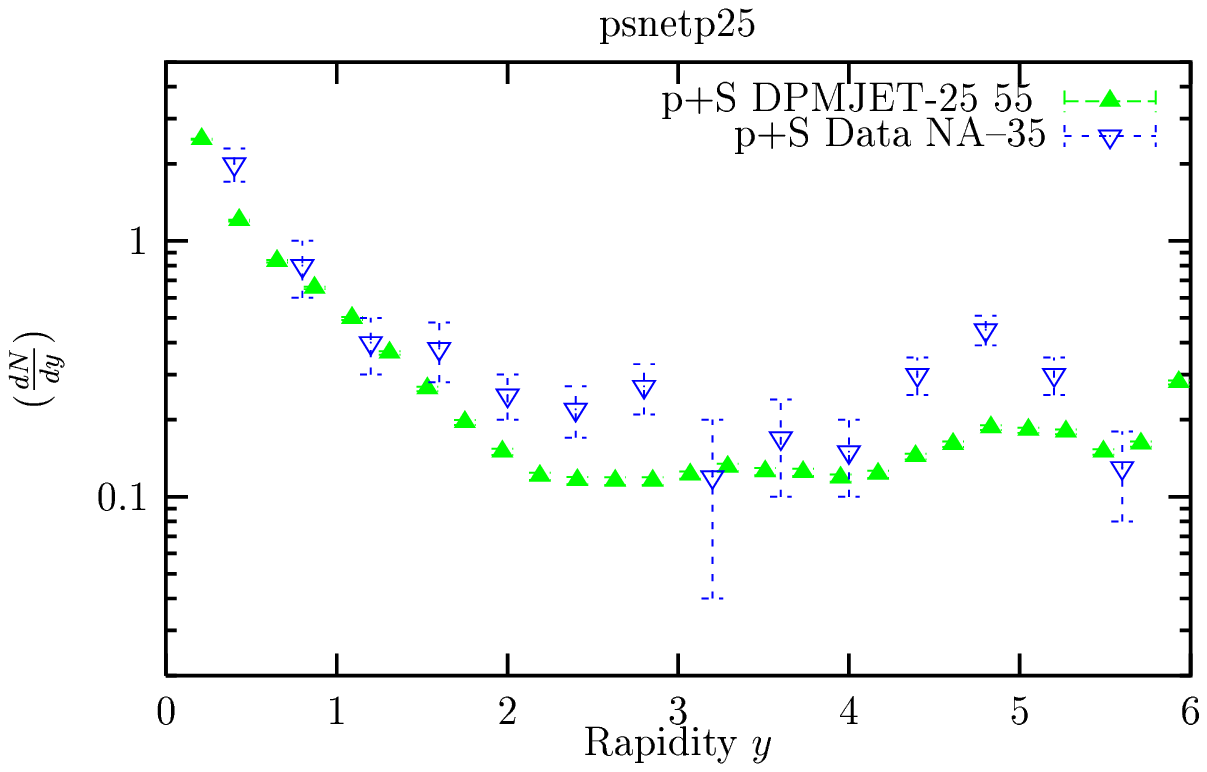}
\end{center}
\begin{center}
 \psfig{figure=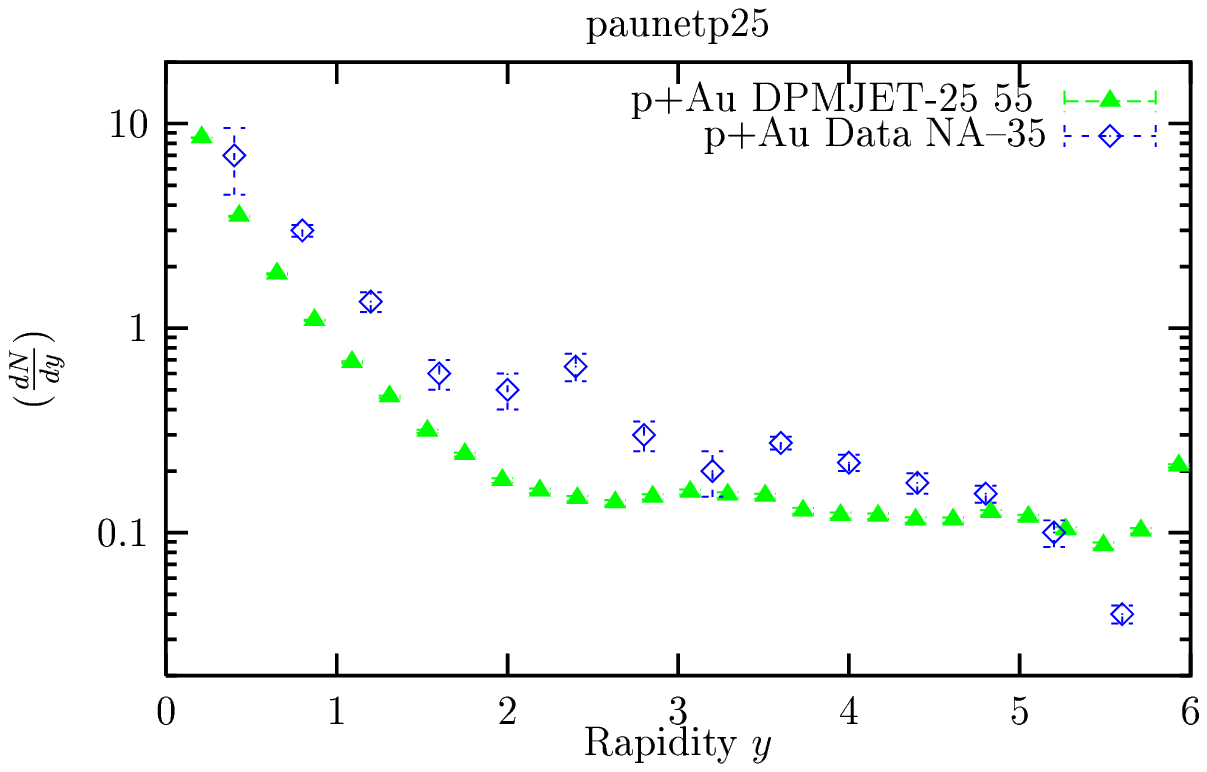}
\end{center}
\vspace*{-3mm}
\caption{{\bf (a) }Net proton ($p-\bar p$) rapidity distribution in p--S
collisions. 
The DPMJET--II.5 results are compared with data
\protect\cite{Alber98}.
{\bf (b) }
Net proton ($p-\bar p$) rapidity distribution in p--Au
collisions. 
The DPMJET--II.5 results are compared with data
\protect\cite{Alber98}.
\label{psnetp25}
}
\end{figure}

In Fig. \ref{psnetp25}.a and \ref{psnetp25}.b we compare the net--proton
distributions according to the full DPMJET--II.5 model with data in p--S
and p--Au collisions \cite{Alber98}. 
Now the dips at central rapidity are filled in the
model, we observe like in the data at central rapidity a flat
net--proton rapidity distribution. 

 \clearpage

\begin{figure}[thb]
\begin{center}
 \psfig{figure=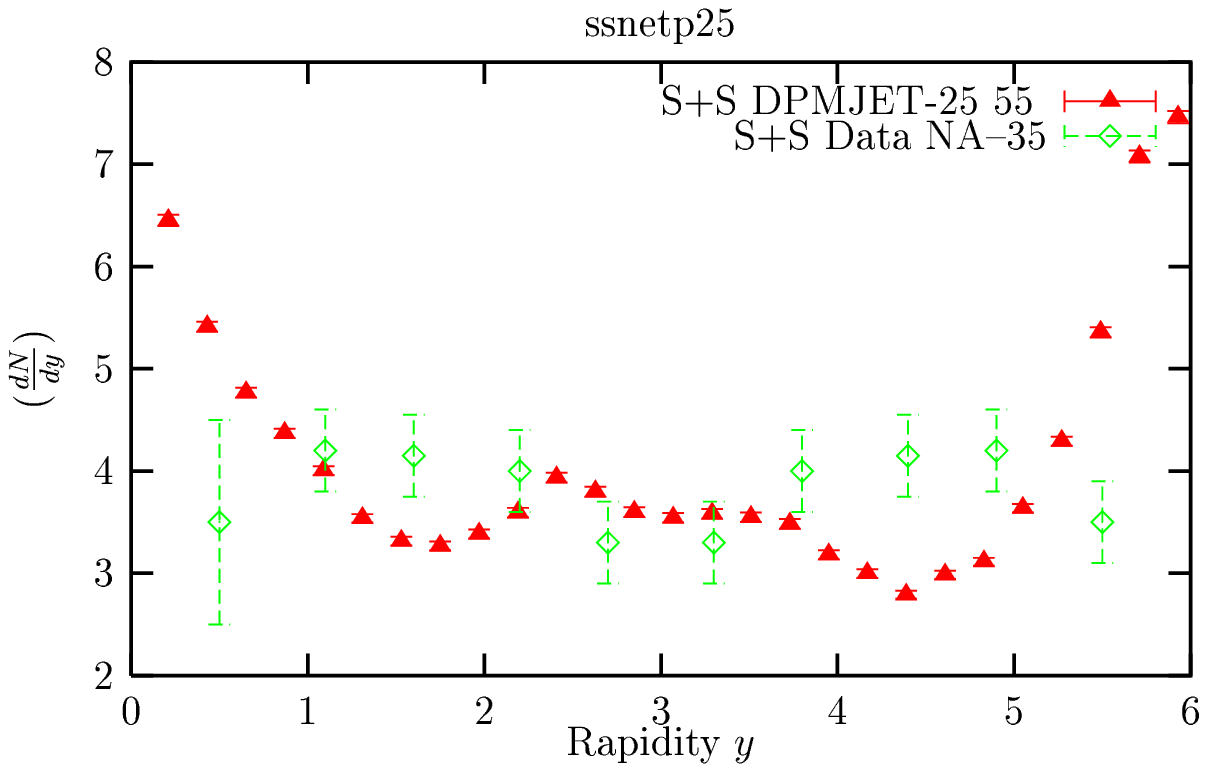}
\end{center}
\begin{center}
 \psfig{figure=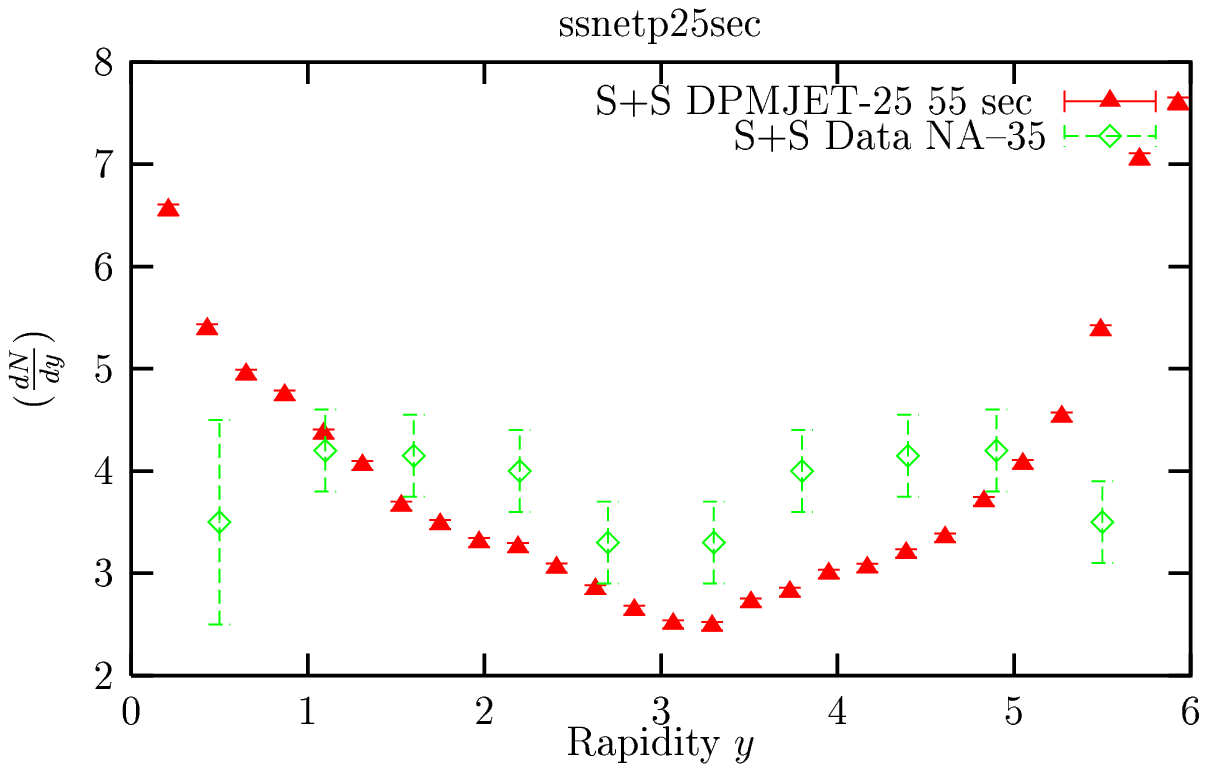}
\end{center}
\vspace*{-3mm}
\caption{Net proton ($p-\bar p$) rapidity distribution in central 
S--S
collisions. 
The DPMJET--II.5 results are compared with data
\protect\cite{Alber98}.
{\bf (a) }DPMJET--II.5 without secondary interactions.
{\bf (b) }DPMJET--II.5 with secondary interactions.
\label{ssnetp25}
}
\end{figure}

In Fig. \ref{ssnetp25} we compare the full DPMJET--II.5 model in the two
versions without and with secondary interactions with data
on net--proton production in central S--S collisions. Also here the dip
at central rapidity in the model 
has disappeared (compare to Fig.\ref{netpnostop}),
however, the agreement
to the data \cite{Alber98} is not perfect. 
There is a significant disagreement in
the fragmentation regions of the two nuclei. The reason for this is in
the spectator evaporation protons included in DPMJET, but apparently not
in the data.

 \clearpage

\begin{figure}[thb]
\begin{center}
 \psfig{figure=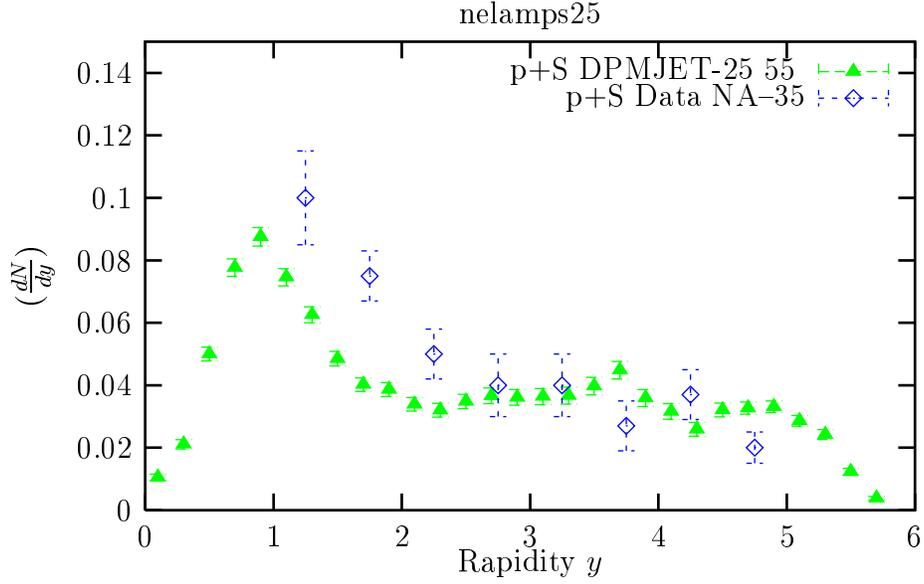}
\end{center}
\vspace*{-3mm}
\caption{Net $\Lambda$ ($\Lambda-\bar \Lambda$) 
rapidity distribution in p--S
collisions. 
The DPMJET--II.5 results are compared with data
\protect\cite{Alber98}.
\label{netlamps25}
}
\end{figure}

\begin{figure}[thb]
\begin{center}
 \psfig{figure=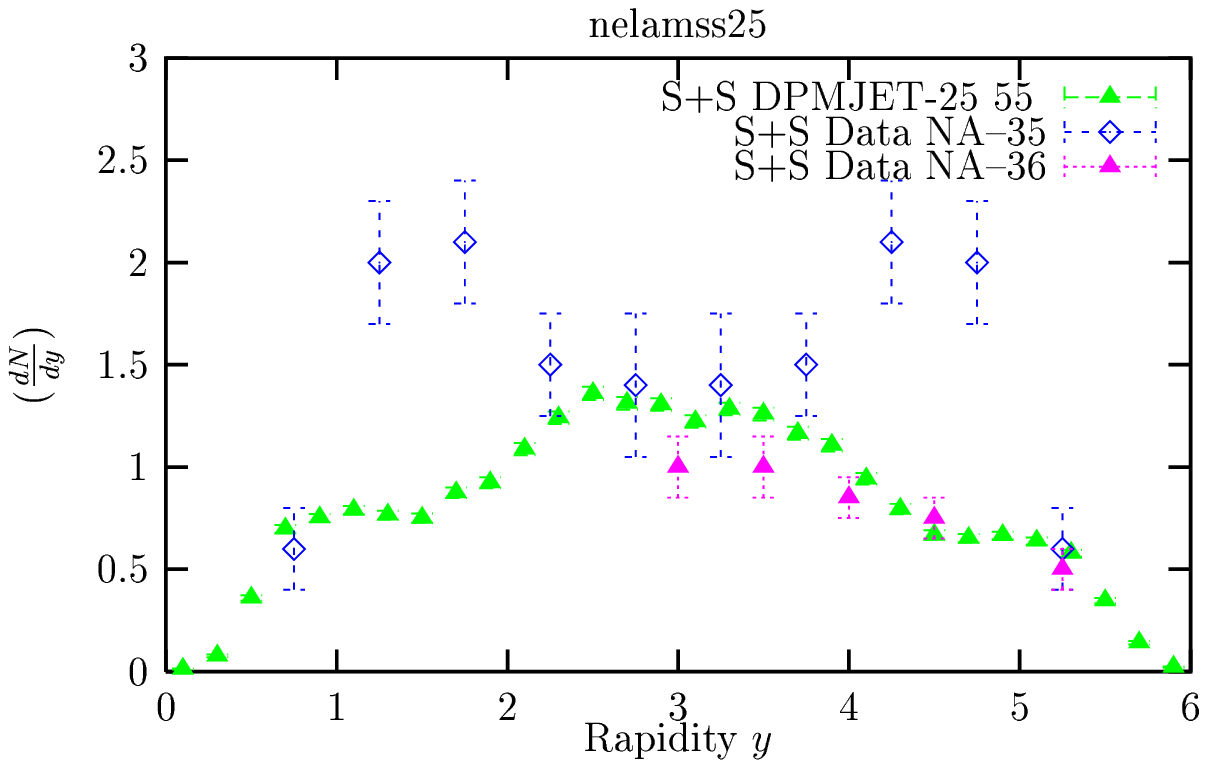}
\end{center}
\begin{center}
 \psfig{figure=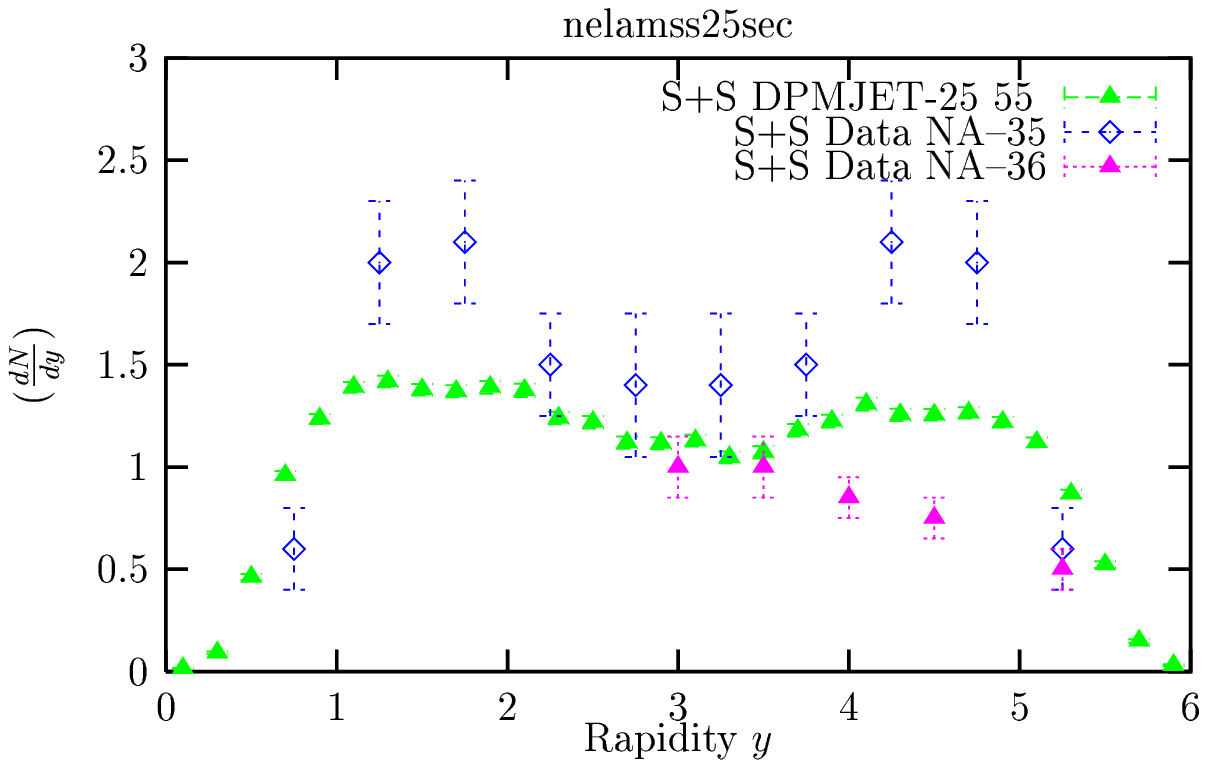}
\end{center}
\vspace*{-3mm}
\caption{
Net $\Lambda$ ($\Lambda-\bar \Lambda$) 
rapidity distribution in central S--S
collisions. 
The DPMJET--II.5 results are compared with data from NA35
\protect\cite{Alber98} and with the total $\Lambda$ production as
measured by NA36 \protect\cite{Judd95}.
{\bf (a) }DPMJET--II.5 without secondary interactions.
{\bf (b) }DPMJET--II.5 with secondary interactions.
\label{netlamss25}
}
\end{figure}

In Fig. \ref{netlamps25}  we compare the full
model to data on net--$\Lambda$ production in p--S 
collisions. In Fig.\ref{netlamps25} the dip at central rapidity has
disappeared in the model. 
In  Fig. \ref{netlamss25}.a and .b we compare the full
model to data on net--$\Lambda$ production  central S--S
collisions for the two models with and without secondary interactions.
 In Fig.\ref{netlamss25}.a and .b again  the central dips  in the models 
 are largely reduced
 but the agreement to the NA35--data is rather
unsatisfactory in the model without secondary interactions.
Also the agreement of the model without secondary interactions
to the total number of
$\Lambda$'s produced according to NA35 is rather unsatisfactory, 
see Table 7.
However, we
find no disagreement to the  total--$\Lambda$ rapidity
distribution as measured by Na36 \cite{Judd95}.

We might conclude from the comparison of net--p and net--$\Lambda$
production, that the secondary interaction is indeed needed to
understand the data, further improvements to the secondary interactions
might be useful.

 \clearpage

 \clearpage
%

\section{Properties of the model in the highest energy region}

In all plots in this Section we present the DPMJET--II.5 results for the
two versions of the model discussed already in Section II.C: 

(i)With only the new diagram for baryon stopping GSQBS, this version is
characterized in the plots by 50, this version corresponds in its high
energy behaviour approximately to the previous version of 
the model DPMJET--II.4.

(ii)With both new diagrams for baryon stopping GSQBS and USQBS, this
version is characterized in the plots by 55, the hight energy
behaviour of this version is new, at high energies the baryons carry 5
to 10 percent less energy than in version 50, correspondingly the
mesons carry more energy than in version 50.

\begin{figure}[thb]
\begin{center}
 \psfig{figure=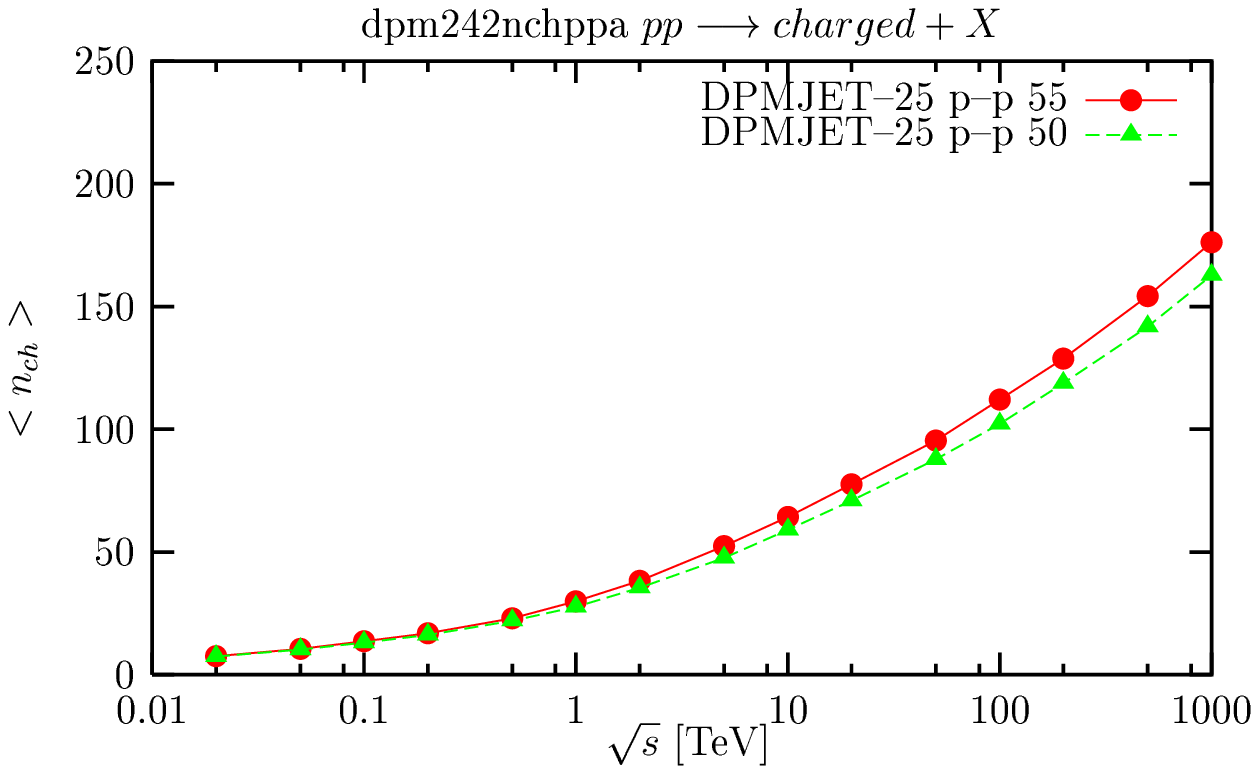}
\end{center}
\begin{center}
 \psfig{figure=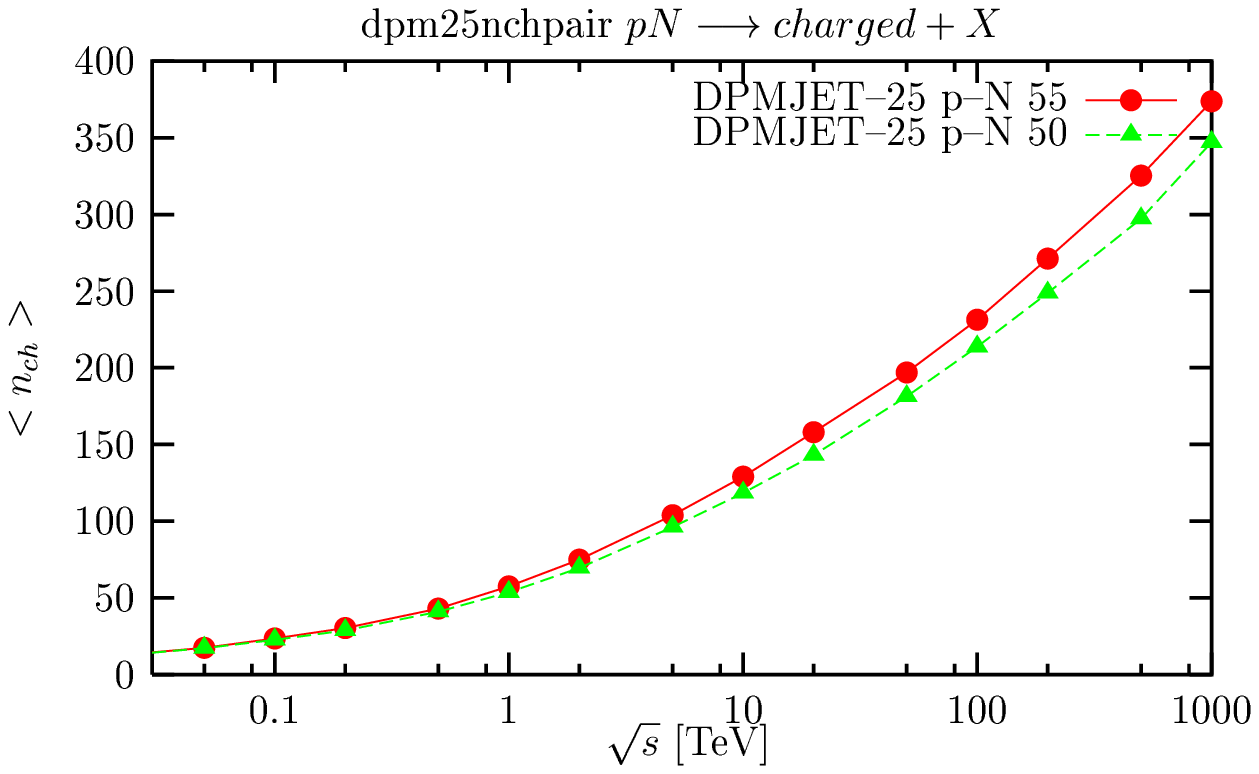}
\end{center}
\vspace*{-3mm}
\caption{Rise of the charged multiplicity in inelastic p--p and p--N
collisions in the center of mass energy range between $\sqrt
s$=0.02 TeV and $\sqrt s$ = 1000 TeV.
\protect\label{dpm222nch}
}
\end{figure}

Fig.\ref{dpm222nch} 
presents the rise of the total charged  multiplicity in p--p and p--N
collisions with the
cms energy $\sqrt s$.
%


The
corresponding rise of the average transverse momenta was already
presented in Fig.\ref{dpm222ptav}

 \clearpage
Following for instance the basic discussion of ~\cite{gaistext},
we introduce a variable $x_{lab}$ similarly to Feynman--$x_F$, but this
time in the lab--frame :

\begin{equation}
x_{lab} = \frac{E_i}{E_0}
\end{equation}

$E_i$ is the lab--energy of a secondary particle $i$ and $E_0$ is the 
lab--energy of the projectile in a h--nucleus collision.
We introduce $x_{lab}$ distributions $F(x_{lab})$ :

\begin{figure}[thb]
\begin{center}
 \psfig{figure=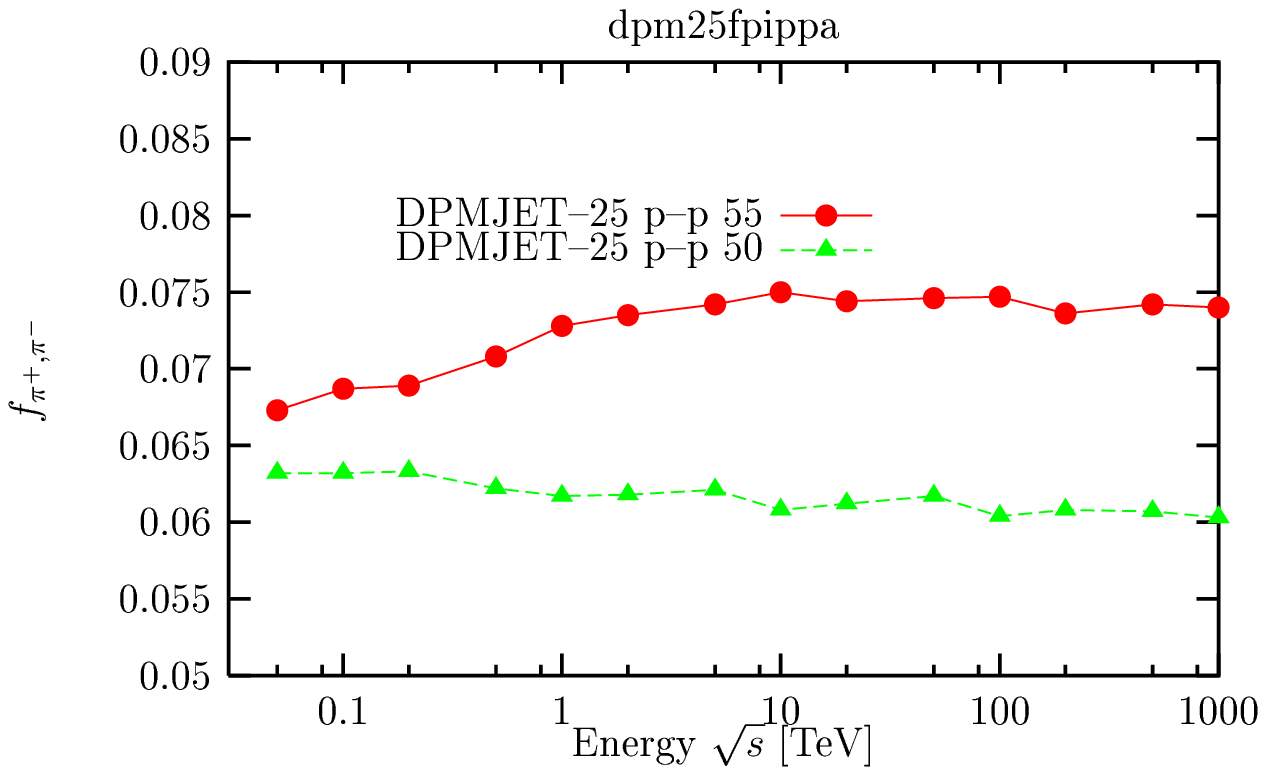}
\end{center}
\begin{center}
 \psfig{figure=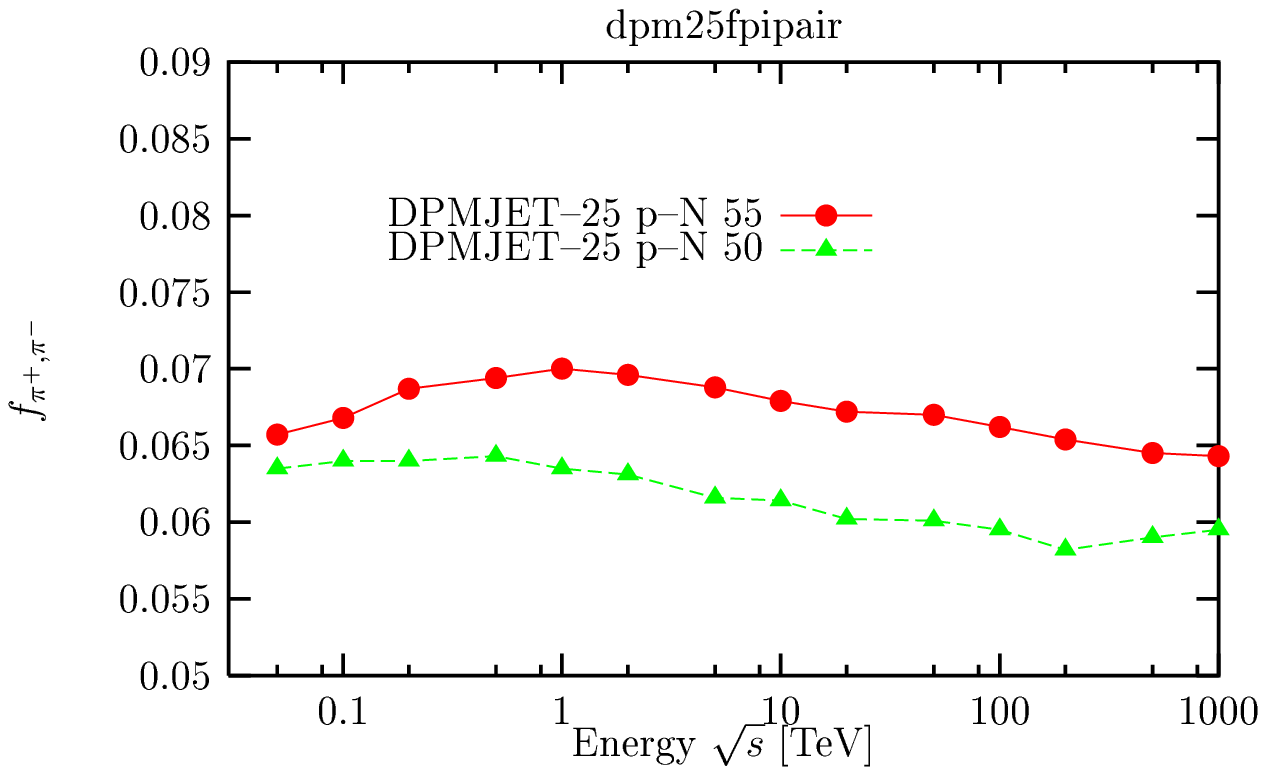}
\end{center}
\vspace*{-3mm}
\caption{Spectrum weighted moments for pion production in p--p and p--N
 collisions as function of the (nucleon--nucleon)
cms energy  $\sqrt s$.
\label{dpm222fpi}
}
\end{figure}

\begin{equation}
F_i(x_{lab}) = x_{lab}\frac{dN_i}{dx_{lab}}
\end{equation}

We note that the Feynman--$x_F$ distribution at positive $x_F$ in the
projectile fragmentation region is a very good approximation to
the $x_{lab}$ distribution. 

 \clearpage
The cosmic ray spectrum--weighted moments in p--A collisions 
are now defined as moments of the $F(x_{lab})$ :
\begin{equation}
f^{p-A}_i = \int^{1}_0 (x_{lab})^{\gamma -1}
F^{p-A}_i(x_{lab})dx_{lab}
\end{equation}
Here $-\gamma \simeq$ --1.7 is the power of the integral cosmic
ray energy spectrum and $A$ represents  the target nucleus.
\begin{figure}[thb]
\begin{center}
 \psfig{figure=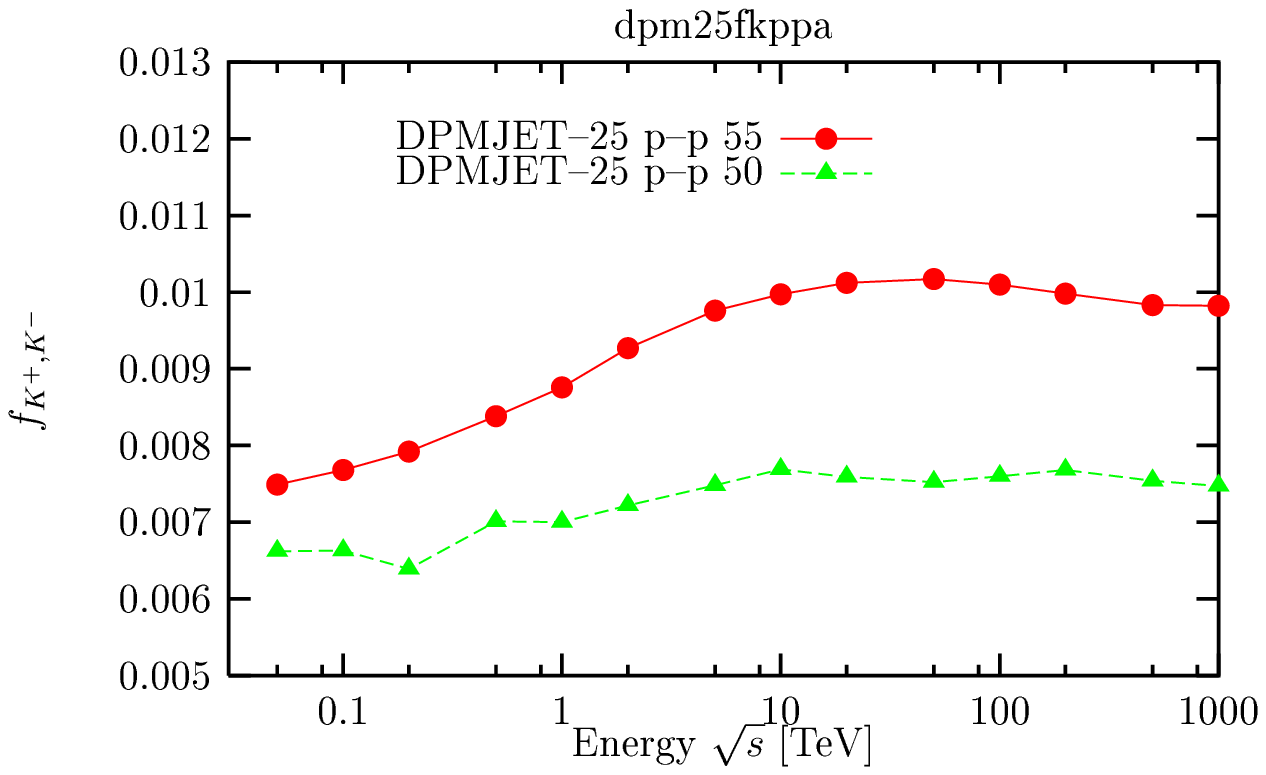}
\end{center}
\begin{center}
 \psfig{figure=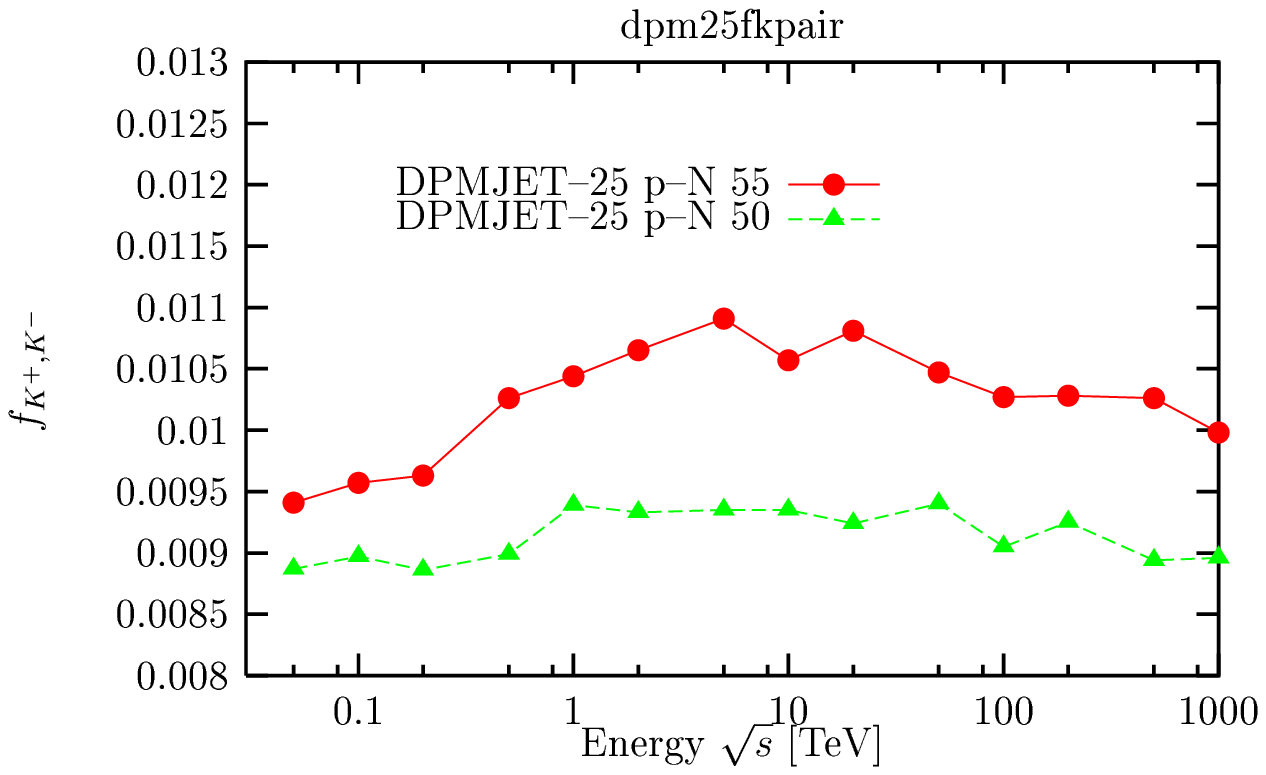}
\end{center}
\vspace*{-3mm}
\caption{Spectrum weighted moments for Kaon production in p--p and p--N
 collisions as function of the (nucleon--nucleon)
cms energy  $\sqrt s$.
\label{dpm222fk}
}
\end{figure}

The spectrum--weighted moments for nucleon--air collisions,
as discussed in ~\cite{gaistext}, determine the
uncorrelated fluxes of energetic particles in the atmosphere.

We also introduce the energy fraction $K^{p-A}_i$ :

\begin{equation}
K^{p-A}_i = \int^{1}_0 
F^{p-A}_i(x_{lab})dx_{lab}
\end{equation}
As for $x_{lab}$, the upper limit for $K$ is 1 in h--nucleus
collisions.

 \clearpage
In Figs.\ref{dpm222fpi}  
we present the spectrum weighted moments
summed over pions of both charges 
 in $pp$ and p--N collisions as function of
the cms energy $\sqrt s$ per nucleon. In Figs.\ref{dpm222fk} 
the
moments are given for charged Kaon production also in $pp$ and
p--N collisions. 

\begin{figure}[thb]
\begin{center}
 \psfig{figure=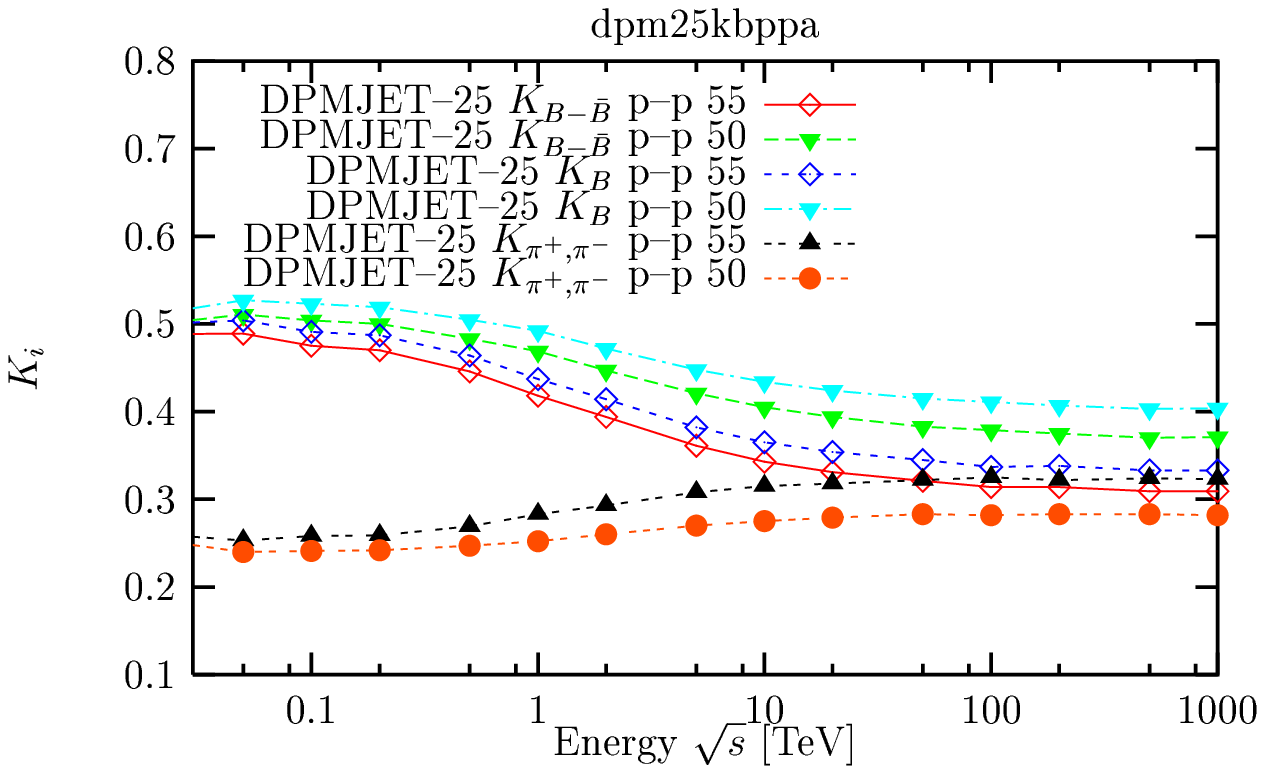}
\end{center}
\begin{center}
 \psfig{figure=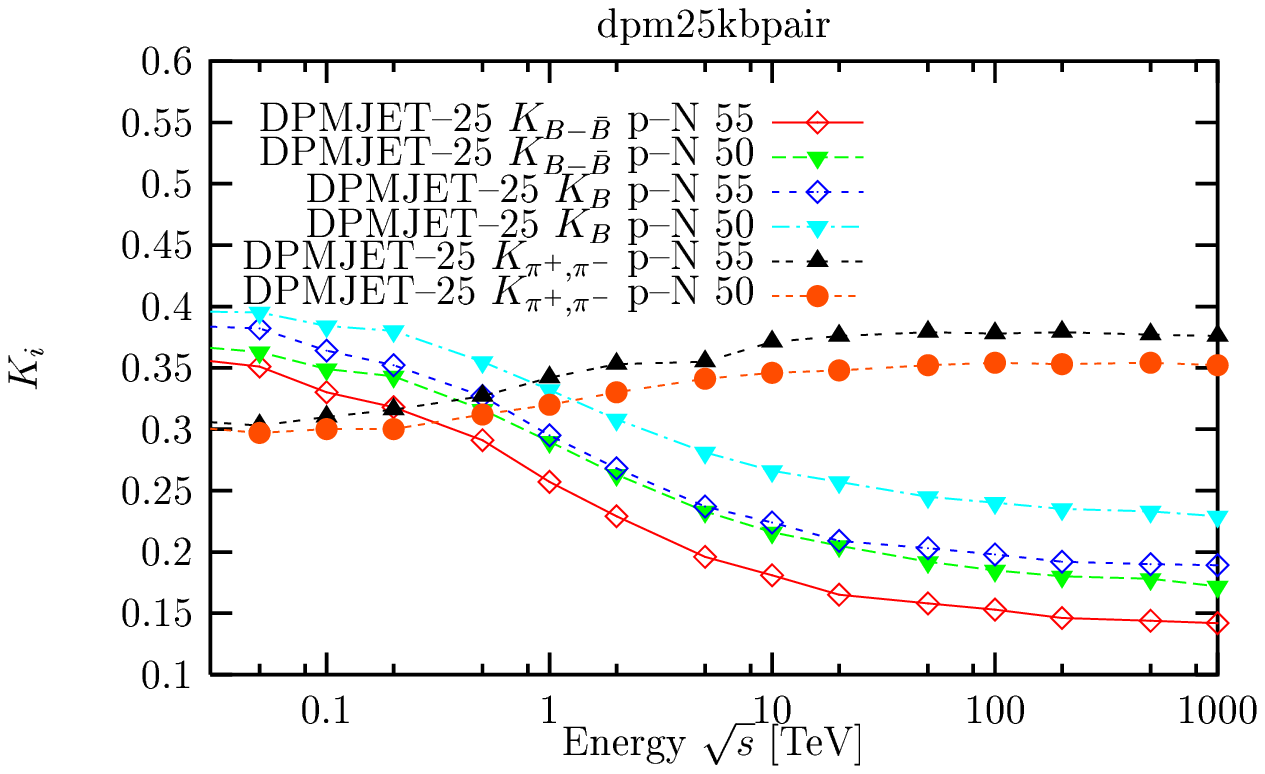}
\end{center}
\vspace*{-3mm}
\caption{Laboratory energy fractions  for net baryons (baryon minus
antibaryon) $B-\bar B$,
baryons  $B$ and
charged pion  production in p--p and p--N
 collisions as function of the (nucleon--nucleon)
cms energy  $\sqrt s$.
\label{dpm222kbpi}
}
\end{figure}
In Fig.\ref{dpm222kbpi}  
we present again for $pp$ anf p--N
collisions the energy fractions K  for net baryons $B-\bar B$ (baryon
minus
antibaryon), B (baryon) and charged pion production. The energy fraction
$K_{B- \bar B}$ is always smaller than the energy fraction $K_B$. The
difference between both is the energy fraction going into antibaryons 
$K_{\bar B}$ (not given separately) which is equal to the energy
fraction carried by the baryons which are produced in baryon--antibaryon
pairs. 

We find in DPMJET--II.5  
all average values characterizing hadron production:
the cross sections (Fig.\ref{xs2126981}), the average transverse
momenta (Fig.\ref{dpm222ptav}) the charged multiplicities 
(Fig.\ref{dpm222nch}), and the moments in Figs. \ref{dpm222fpi},
\ref{dpm222fk} and \ref{dpm222kbpi} to change smoothly with
energy in most cases just like the logarithm of the energy.

So far, we have not found any experimental 
data to favour either the versions of DPMJET--II.5
(i or 50) with only GSQBS or (ii or 55) with GSQBS
and USQBS.  
 Clearly, the version prefered by theoretical prejudices is
(ii or 55). In this version we have a better Feynman scaling of meson
distibutions and spectrum weighted moments and we have a faster decrease
 with the collision energy of the energy fractions into secondary
 baryons.

\clearpage

%
%
\section{Summary}

In the present paper we present new features of the DPMJET--II.5 model
and  the properties of the model in the Cosmic Ray energy region.
The most important of the new features are the new GSQBS and USQBS
diagrams, which modify significantly the baryon stopping in the model.
Using as well the GSQBS and USQBS options we find significant
differences in the high energy behaviour 
of the new model against former versions of DPMJET.

We  report  also
about comparisons of DPMJET--II.5
 with  accelerator and collider experiments.
 We find especially an improved agreement of the Feynman $x$
 distributions of leading protons to data and a reasonable agreement to
 experimental data on leading baryon stopping in nuclear collisions.
Many of the published comparisons of former versions of the
model to other aspects of experimental data  come out
quite similar with   DPMJET--II.5, we mention
especially 
 the
properties of the model at low energies
\cite{Ranftsare95,Ferrari95a,Ferrari96a} and charm production
\cite{dpmcharm}.

 There
are certainly more tests of the model needed  and new features to be
incorporated for hadron--nucleus
and nucleus--nucleus collisions,
 for strangeness production and for secondary
interactions of the produced particles. 
DPMJET will continue to evolve, there are already plans for a major new
update to DPMJET--III.
%
\vspace{3cm}
\section*{Acknowledgements}
 I thank first of all my  collaborator Dr. H.--J. M\"ohring
 with whom together  previous versions of the code up to
 DTUNUC--I.03 were developed. 
The author  thanks Y.~Shmakov for supplying the DIAGEN code prior
to publication. Furthermore,  the support
of  CERN , the Department of Theoretical Physics in Lund, INFN,
Sezione
di Milano ,
INFN,LNF Frascati, LAPP Annecy,
The University Santiago de Compostela and INFN, Lab. Naz. del
Gran Sasso and Siegen University
where parts of the program were developped is acknowledged. The
code in the version  described here was finalized  
at Siegen. 
The author acknowledges   
the fruitful collaborations with P.Aurenche, G.Battistoni, 
F.Bopp, M.Braun, A.~Capella,
R.Engel, A.Ferrari, C.Forti, K.H\"an\ss gen, K.Hahn, 
I.Kawrakov, C.Merino, N.Mokhov, H.J.M\"ohring, C.Pajares,
D.Pertermann, S.Ritter, S.Roesler, P.Sala  and
J.Tran~Thanh~Van on the Dual Parton Model in general and he thanks
G.Battistoni,  F.Bopp, R.Engel and S.Roesler 
for useful comments on this paper. 

%
%
 \newpage
%
\bibliographystyle{zpc}
\bibliography{dpm11}
 
 \clearpage

%
 \end{document}